\documentclass[nonacm, acmsmall]{acmart}

\usepackage{listings}                           
\usepackage{float}
\usepackage{subcaption}                          
\usepackage{siunitx} 
\usepackage{graphicx}
\usepackage{svg}
\svgsetup{inkscapelatex=true} 

\usepackage{tcolorbox}
\usepackage{tikz}
\usepackage{tabularx}            
\usepackage{array}              
\usepackage{booktabs}            
\usepackage{xcolor, colortbl}    
\usepackage{makecell}

\usepackage[font=small,skip=2pt]{caption}

\definecolor{lightgray}{gray}{0.85} 

\usepackage{booktabs}
\usepackage{multirow}
\usepackage{enumitem}
\usepackage{pifont}

\newcommand{\cmark}{\ding{51}}
\newcommand{\xmark}{\ding{55}}

\newfloat{Listing}{tbp}{code}
\floatname{Listing}{Listing}
\DeclareCaptionSubType{Listing}
\newcommand{\name}{Code\-Cure\-Agent}

\definecolor{green-blue}{HTML}{2D7D74}  
\definecolor{dark-green}{HTML}{2E7D32}
\definecolor{light-gray}{gray}{0.95}

\definecolor{pblue}{rgb}{0.13,0.13,1}
\definecolor{pgreen}{rgb}{0,0.5,0}
\definecolor{pbrown}{rgb}{0.6,0,0}
\definecolor{pgrey}{rgb}{0.46,0.45,0.48}

\definecolor{lightgray}{HTML}{f6f6f6}
\definecolor{darkgray}{rgb}{.4,.4,.4}
\definecolor{darkblue}{HTML}{1b4db3}
\definecolor{brickred}{HTML}{b04f4f}
\definecolor{purple}{rgb}{0.65, 0.12, 0.82}
\definecolor{diffadd}{HTML}{288f26}
\definecolor{diffrmbg}{HTML}{ffebe9}
\definecolor{diffaddbg}{HTML}{e6ffeb}
\definecolor{diffremove}{HTML}{de4f54}
\definecolor{carrotorange}{rgb}{0.8, 0.33, 0.0}
\definecolor{highlight}{HTML}{fefbc2}

\definecolor{bluegray}{HTML}{3182bd}
\definecolor{delim}{RGB}{20,105,176}
\definecolor{numb}{RGB}{106, 109, 32}
\definecolor{string}{rgb}{0.64,0.08,0.08}

\definecolor{Gray}{gray}{0.6}
\definecolor{LightGray}{gray}{0.9}
\definecolor{bananayellow}{rgb}{1.0, 0.88, 0.5}
\definecolor{myred}{rgb}{1.0,0.44,0.37}

\newcommand{\code}[1]{\texttt{\small#1}} 
\newcommand{\scode}[1]{\texttt{\footnotesize#1}} 

\lstdefinelanguage{Java}{
	keywords={new, true, false, catch, void, public, return, null, catch, int},
	keywordstyle=\color{darkblue}\bfseries,
	ndkeywords={class, export, boolean, throw, implements, import, this, setTimeout},
	ndkeywordstyle=\color{brickred}\bfseries,
	identifierstyle=\color{black},
	sensitive=false,
	comment=[l]{//},
	morecomment=[f][\color{diffadd}\bfseries]{+\ },
	morecomment=[s]{/*}{*/},
	morecomment=[f][\color{diffremove}\bfseries]{- },
	commentstyle=\color{violet}\ttfamily,
	stringstyle=\color{carrotorange}\ttfamily,
	morestring=[b]',
	morestring=[b]"
}

\lstdefinelanguage{diffWithComment}{
  morecomment=[l][\color{diffremove}\bfseries]{-},
  morecomment=[l][\color{diffadd}\bfseries]{+},
  morecomment=[l][\color{gray}\bfseries]{@@},
  morecomment=[l]{//},
  literate=
    {−}{{-}}1 
    {–}{{-}}1 
    {—}{{-}}1 
}

\lstdefinelanguage{diff}{
  morecomment=[l][\color{diffremove}\bfseries]{-},
  morecomment=[l][\color{diffadd}\bfseries]{+},
  morecomment=[l][\color{gray}\bfseries]{@@},
  literate=
    {−}{{-}}1 
    {–}{{-}}1 
    {—}{{-}}1 
}

\definecolor{jsonbg}{HTML}{F4F6F8}
\definecolor{jsonframe}{gray}{0.80}
\definecolor{jsontext}{HTML}{111111}
\definecolor{jsonkey}{HTML}{0B3D91}      
\definecolor{jsonstring}{HTML}{1B5E20}   
\definecolor{jsonnumb}{HTML}{AA2E25}     
\definecolor{jsonbool}{HTML}{7F0055}     
\definecolor{jsondelim}{gray}{0.25}      
\definecolor{jsoncomment}{gray}{0.45}

\usepackage{listings}
\lstdefinelanguage{json}{
	basicstyle=\ttfamily\scriptsize\color{jsontext},
	numbers=none,
	numberstyle=none,
	numbersep=none,
	frame=single,
	rulecolor=\color{jsonframe},
	backgroundcolor=\color{jsonbg},
	showstringspaces=false,
	showspaces=false,
	showtabs=false,
	columns=fullflexible,
	upquote=true,
	breaklines=true,
	breakatwhitespace=true,
	postbreak=\raisebox{0ex}[0ex][0ex]{\ensuremath{\color{gray}\hookrightarrow\space}},
	morestring=[b]",
	stringstyle=\color{jsonstring},
	morekeywords={true,false,null},
	keywordstyle=\color{jsonbool}\bfseries,
	moredelim=**[is][\color{jsonkey}\bfseries]{\{"}{":},
	moredelim=**[is][\color{jsonkey}\bfseries]{,"}{":},
	literate=
	*{:}{{{\color{jsondelim}{:}}}}1
	{,}{{{\color{jsondelim}{,}}}}1
	{]}{{{\color{jsondelim}{]}}}}1
	{0}{{{\color{jsonnumb}0}}}1
	{1}{{{\color{jsonnumb}1}}}1
	{2}{{{\color{jsonnumb}2}}}1
	{3}{{{\color{jsonnumb}3}}}1
	{4}{{{\color{jsonnumb}4}}}1
	{5}{{{\color{jsonnumb}5}}}1
	{6}{{{\color{jsonnumb}6}}}1
	{7}{{{\color{jsonnumb}7}}}1
	{8}{{{\color{jsonnumb}8}}}1
	{9}{{{\color{jsonnumb}9}}}1,
	}

\lstdefinestyle{jsoncompact}{
language=json,
aboveskip=4pt, belowskip=4pt,
xleftmargin=1.0em, framexleftmargin=0.8em, framesep=3pt
}

\lstdefinestyle{JavaCodeStyle}{
	language=Java,
	backgroundcolor=\color{lightgray},
	extendedchars=true,
	basicstyle=\scriptsize\ttfamily,
	escapeinside={(*@}{@*)},
	showstringspaces=false,
	showspaces=false,
	numbers=left,
	numberstyle=\scriptsize,
	numbersep=6pt,
	tabsize=4,
	breaklines=true,
	showtabs=false,
	captionpos=b,
	frame=single,
	framesep=4pt,
	linewidth=.98\columnwidth,
	xleftmargin=15pt,        
	rulecolor=\color{lightgray},        
	literate={-}{{{\color{black}-}}}1,
}
\lstdefinestyle{JavaInlineStyle}{
	language=Java,
	backgroundcolor=\color{lightgray},
	extendedchars=true,
	basicstyle=\ttfamily,
	escapeinside={(*@}{@*)},
	showstringspaces=false,
	showspaces=false,
	numbers=left,
	numberstyle=\scriptsize,
	numbersep=6pt,
	tabsize=4,
	breaklines=true,
	showtabs=false,
	captionpos=b,
	frame=single,
	framesep=4pt,
	linewidth=.98\columnwidth,
	xleftmargin=10pt,        
	rulecolor=\color{lightgray},        
	literate={-}{{{\color{black}-}}}1,
}

\lstdefinestyle{DiffCodeStyle}{
	language=diff,
	backgroundcolor=\color{lightgray},
	extendedchars=true,
	basicstyle=\scriptsize\ttfamily,
	escapeinside={(*@}{@*)},
	showstringspaces=false,
	showspaces=false,
	numbers=left,
	numberstyle=\scriptsize,
	numbersep=6pt,
	tabsize=4,
	breaklines=true,
	showtabs=false,
	captionpos=b,
	frame=single,
	framesep=4pt,
	linewidth=.98\columnwidth,
	xleftmargin=15pt,        
	rulecolor=\color{lightgray}
}

\lstdefinestyle{DiffWithCommentCodeStyle}{
	language=diffWithComment,
	backgroundcolor=\color{lightgray},
	extendedchars=true,
	basicstyle=\scriptsize\ttfamily,
	escapeinside={(*@}{@*)},
	showstringspaces=false,
	showspaces=false,
	numbers=left,
	numberstyle=\scriptsize,
	numbersep=6pt,
	tabsize=4,
	breaklines=true,
	showtabs=false,
	captionpos=b,
	frame=single,
	framesep=4pt,
	linewidth=.98\columnwidth,
	xleftmargin=15pt,        
	rulecolor=\color{lightgray}
}

\begin{document}

\title{\name{}: Automatic Classification and Repair of Static Analysis Warnings}

\author{Pascal Joos}
\affiliation{%
  \institution{CISPA Helmholtz Center for Information Security}
  \country{Germany}}
\email{pascal.joos@outlook.de}

\author{Islem Bouzenia}
\affiliation{%
  \institution{CISPA Helmholtz Center for Information Security}
  \country{Germany}}
\email{bouzenia.islem@pm.me}

\author{Michael Pradel}
\affiliation{%
  \institution{CISPA Helmholtz Center for Information Security}
  \country{Germany}}
\email{michael@binaervarianz.de}

\begin{abstract}
Static analysis tools are widely used to detect bugs, vulnerabilities, and code smells.
Traditionally, developers must resolve these warnings manually by analyzing the warning, deciding whether to fix or suppress it, and validating the correctness of the code change.
Because this process is tedious, developers sometimes ignore warnings, leading to an accumulation of warnings and a degradation of code quality.
Prior work suggests techniques to automatically repair static analysis warnings, but these are limited to specific analysis rules, cannot perform multi-file edits, and rely on weak validation mechanisms.
This paper presents \name{}, an approach that harnesses LLM-based agents to automatically analyze, classify, and repair static analysis warnings.
Unlike previous work, our method does not follow a predetermined algorithm.
Instead, we adopt an agentic framework that iteratively invokes tools to gather additional information from the codebase (e.g., via code search) and edit the codebase to resolve the warning.
\name{} detects and suppresses false positives, while fixing true positives when identified.
We equip \name{} with a three-step heuristic to approve patches: (1) build the project, (2) verify that the warning disappears without introducing new warnings, and (3) run the test suite.
We evaluate \name{} on a dataset of \num{1000} SonarQube warnings found in 106 Java projects and covering 291 distinct rules.
Our approach produces plausible fixes for 96.8\% of the warnings, outperforming state-of-the-art baseline approaches by 29.2\%--34.0\% in plausible-fix rate. 
Manual inspection of 291 cases reveals a correct-fix rate of 86.3\%, showing that \name{} can reliably repair static analysis warnings.
The approach incurs LLM costs of about 2.9 cents (USD) and an end-to-end processing time of about four minutes per warning.
We envision \name{} helping to clean existing codebases and being integrated into CI/CD pipelines to prevent the accumulation of static analysis warnings.

\end{abstract}

\maketitle

\section{Introduction}
Static analysis tools help developers identify potential bugs, vulnerabilities, and code smells~\cite{Youn2023,Le2022a,issta2023-Fluffy}, contributing to clean, reliable, secure, and maintainable software.
Such tools check a code base for violations of a predefined set of rules, and report warnings when a rule is violated.
Once a warning is reported, the task of resolving it traditionally falls on the developers.

Manually handling static analysis warnings poses several challenges.
First, developers must understand the violated rule and why it is reported.
Second, they must decide whether the warning is a true positive that warrants a fix, or a false positive that can be safely suppressed~\cite{fse2025_suppressions}.
Third, developers must devise an appropriate fix and implement it, which may range from a quick, local edit to substantially larger changes that span multiple files.
Fourth, any new warnings introduced by the fix should undergo the same process. 
Finally, to be confident that the fix does not break functionality, the code change must be validated, e.g., by executing all relevant tests.
This labor-intensive and repetitive workflow limits the practical adoption of static analyzers, causing unresolved warnings to accumulate in software projects.
Moreover, introducing a static analyzer into an active project incurs substantial effort to address existing warnings, which can disrupt productivity and pollute build and testing pipelines with warning reports.

To assist developers in this process, automated classification and repair techniques have been proposed.
Some studies focus on classifying warnings as true positives or false positives~\cite{Kharkar2022,Wen2024,Li2024}, while others aim to automatically fix warnings~\cite{Etemadi2023a,icse2024-PyTy,Jain2023,Liu2023}.
In the literature, two families of approaches prevail: rule-based heuristics~\cite{Etemadi2023a,Bavishi2019,oopsla2019} and learning-based methods~\cite{icse2024-PyTy,Chen2023a,Dilhara2024}.
Rule-based approaches target a limited set of warning types and implement heuristic transformations for them, which is fast but covers only a few rules.
A representative work in this category is Sorald~\cite{Etemadi2023a}, which implements rule-based fixes for 30 SonarQube rules (out of 479 default rules).
Learning-based approaches have become increasingly popular, especially with the success of large language models (LLMs).
For example, PyTy~\cite{Chow_2024} trains a language model to repair static type errors in Python. Likewise, iSMELL~\cite{Wu2024iSMELL} and CORE~\cite{Wadhwa2024CORE:LLMs} query an LLM to automatically fix warnings.

Despite these advances, current approaches still suffer from several fundamental and practical limitations. To begin with, existing methods are confined to a single function or file scope, whereas some warnings require modifications across multiple files.
Moreover, current repair techniques treat all warnings as true positives without verifying whether they should be fixed~\cite{Etemadi2023a,Jain2023,Wen2024,Kharkar2022}.
This ignores the fact that static analyzers emit false positives, which developers would simply suppress~\cite{fse2025_suppressions}, instead of devising a ``fix'' for a non-existent problem.
Additionally, current approaches~\cite{Etemadi2023Sorald:Violations,Wadhwa2024CORE:LLMs} lack robust validation of the suggested fixes, as they are not subjected to systematic testing.
Finally, LLM-based methods inherit the usual drawbacks of language models, namely outdated training data~\cite{codexStudy2022,Li2023a} and hallucination~\cite{Gu2024,Chen2023a}.

This paper presents \name{}, an end-to-end approach for classifying and fixing static analysis warnings.
\name{} follows an agentic paradigm, leveraging an LLM in a fully autonomous loop, where an agent explores the project and makes edits via tools similar to those used by human developers.
The agentic architecture allows \name{} to retrieve relevant documentation and learn from previous attempts, alleviating typical LLM limitations such as outdated knowledge or hallucination.
The agent is divided into two sub-agents.
The first sub-agent determines whether a warning is a true positive or a false positive, thereby avoiding the unfounded assumption that every warning requires a fix.
Depending on this classification, the second sub-agent either repairs the warning (true positive) or suppresses it (false positive).
Unlike prior work, \name{} can modify multiple files and multiple locations within a file, enabling it to address complex warnings more effectively.
A key component of \name{} is the \textit{change approver}, which builds the project, confirms that the target warning disappears without introducing new warnings, and runs the test suite to identify and reject low-quality patches.
We evaluate \name{} on a dataset of \num{1000} SonarQube warnings extracted from 106 Java projects. Running \name{} on this dataset produces plausible fixes for 968 warnings (96.8\%), with an average processing time of four minutes per warning.
Overall, \name{} classifies 69.6\% of the warnings as true positives and 30.4\% as false positives, highlighting the prevalence of false positives.
Manual inspection of 291 warnings with distinct rules shows that 91.8\% of the classifications are correct, 89.0\% of the fixes are plausible, and 86.3\% of the fixes are correct according to our manual validation.
\name{} handles warnings of varying difficulty:
Of the 968 plausible fixes, 424 modify multiple lines, and 27 affect multiple files.
Under the same evaluation process, Sorald~\cite{Etemadi2023Sorald:Violations} can process 62/\num{1000} warnings and produces plausible fixes for 69.4\% of them, whereas \name{} produces plausible fixes for all 62 (+30.6\%).
iSMELL~\cite{Wu2024iSMELL} and CORE~\cite{Wadhwa2024CORE:LLMs} yield plausible fixes for 62.8\% and 67.6\% of the \num{1000} warnings, respectively, which is 34.0\% and 29.2\% less than \name{}.

Given its novel agentic design, an end-to-end pipeline (classification, fixing, and validation), and low operational cost, \name{} offers effective automatic repair of static analysis warnings. We anticipate that our results will encourage developers to adopt static analyzers without fearing excessive time or monetary overhead.

In summary, our contributions are:

\begin{itemize} 
	
	\item The first LLM agent for addressing static analysis warnings, which autonomously explores the codebase and applies edits via tools.
	
	\item A three-stage design that (i) classifies warnings as true positives or false positives, and (ii) either fixes or suppresses them, and (iii) validates the changes via building, re-analyzing, and testing.
	
	\item A comprehensive evaluation on a dataset of \num{1000} warnings from 106 real-world projects, demonstrating that \name{} outperforms state-of-the-art baselines by a large margin.
	
	\item The code and data are publicly available at \href{https://github.com/sola-st/CodeCureAgent}{https://github.com/sola-st/CodeCureAgent}.
	
\end{itemize}

\section{Motivation and Running Example}

Before presenting the details of \name{}, the following motivates our work by discussing key challenges in fixing static analysis warnings and illustrating them with real-world examples.

\subsection{Challenge 1: Not All Warnings Need to be Fixed}

Static analyzers are designed to identify issues in code.
Unfortunately, not all warnings reported by static analysis are useful to developers, e.g., because some warnings are false positives or the developers do not consider them important enough to address~\cite{Rutar2004,johnson2013don,imtiaz2019challenges}. 
Instead, developers often choose to suppress warnings, e.g., by adding annotations or comments to the code~\cite{fse2025_suppressions}.

\begin{Listing}
	\begin{lstlisting}[style=JavaCodeStyle]
import lombok.Value;

public class PlayerEnchantOptions {
	// Annotation to auto-generate constructor, getters, setters
	@Value
	public class EnchantOptionData {
		private final String enchantName; // Raises SonarQube warning S1068:
		                                  // "Unused 'private' fields should be removed"
	}
	
	public void addEnchantOption(String enchantName){
		this.options.add(
			// Call to auto-generated constructor
		    new EnchantOptionData(..., enchantName, ...)
...
	\end{lstlisting}
	\caption{Simplified example of false positive reported in project \href{https://github.com/CloudburstMC/Nukkit/blob/47d54ad26109b5d89936e0d9123c591a99a2cf60/src/main/java/cn/nukkit/network/protocol/PlayerEnchantOptionsPacket.java\#L90}{Nukkit}.}
	\label{l:FalsePositiveLombokExample}
\end{Listing}

Listing~\ref{l:FalsePositiveLombokExample} illustrates an example of a false positive, which when fixed in the way suggested by the analysis documentation, will break the code's functionality.
The warning indicates that the \code{enchantName} field is private but never used in the scope of the class \code{EnchantOptionData}. The documentation associated with the analysis rule instructs the developer to remove the unused private field from the class.
While this change seems reasonable at first glance, the class turns out to use an \code{@Value} annotation from a third-party library (Lombok) that automatically generates constructors, getters, and other boilerplate methods.
Interestingly, a later version of the static analyzer (SonarQube v10.6) acknowledges this false positive and adds special treatment for it.

Prior work on automatically fixing static analysis warnings~\cite{Etemadi2023Sorald:Violations,Wadhwa2024CORE:LLMs,icse2024-PyTy} ignores the possibility of false positives and assumes that all warnings need to be fixed.
Instead, our approach addresses this challenge by first classifying whether a warning is a true positive or a false positive and then either fixing or suppressing the warning, respectively.

\subsection{Challenge 2: Choosing the Best Fix for a Warning May Require Non-Local Context}

\begin{Listing}
	\begin{lstlisting}[style=JavaCodeStyle]
public class ChangeInfo {
	public int _number; // Raises SonarQube warning S1104: 
	                    // "Class variable fields should not have public accessibility"
	...
}
	\end{lstlisting}
	\caption{Running example. Simplified code with warning of rule S1104 in project \href{https://github.com/uwolfer/gerrit-rest-java-client/blob/de49b4e19572334d60f381955b72e1aa8bbbab49/src/main/java/com/google/gerrit/extensions/common/ChangeInfo.java\#L82}{gerrit-rest-java-client}.}
	\label{l:RunningExample}
\end{Listing}

Once a warning has been determined to be worth fixing, the next challenge is to find an appropriate fix.
For many warnings, there are multiple possible fixes, and choosing the best one may require understanding the code context beyond the warnings location.
To illustrate this challenge, consider our running example in Listing~\ref{l:RunningExample}.
The analysis warns about the public accessibility of the field \code{\_number} in the class \code{ChangeInfo}.
The documentation associated with the violated analysis rule describes two options for fixing the warning: (1) make the field private and provide getter and setter methods, or (2) mark the field as \code{public final}.
Deciding which option is most appropriate requires understanding the usage of the field beyond the lines surrounding the warning:
Option (1) is suitable if instances of the class may have different values for the field and the value may change, while option (2) is only applicable if the field is intended to never change its value.

One potential approach for addressing this challenge is to provide relevant code snippets to the LLM.
However, it is not trivial to determine which parts of the code are relevant, especially when aiming to support a wide range of warning types.
Instead of hard-coding heuristics for selecting relevant code snippets, \name{} addresses this challenge by equipping an LLM agent with tools to actively gather code and other information relevant to a warning, e.g., by searching for references to the field in other parts of the codebase or by looking up documentation.

\subsection{Challenge 3: Fixes May Span Multiple Hunks and Multiple Files}

While some warnings can be fixed by modifying a few lines of code around the warning location, other warnings require more extensive changes that may span multiple hunks and multiple files.
This poses a challenge for automated repair approaches, especially those that operate on a single hunk~\cite{icse2024-PyTy} or single file~\cite{Wadhwa2024CORE:LLMs}.
For the running example in Listing~\ref{l:RunningExample}, the appropriate fix is to make the \code{\_number} field private and provide getter and setter methods.
Because the field is accessed by other classes, the fix also requires modifying all references to the field outside of the class by replacing any reference to the \code{\_number} field with a call to the getter or setter method.
That is, the complete fix spans multiple hunks and multiple files.

Our approach addresses this challenge in two ways.
First, we provide the LLM agent with a tool that can modify multiple locations in multiple files in a single step.
Second, the agent may invoke this tool multiple times, thereby allowing it to iteratively refine its fix.

\subsection{Challenge 4: Candidate Fixes May Break Functionality or Introduce New Warnings}

A final challenge faced by a developer or an automated repair approach is to ensure that the proposed fix does not break the functionality of the code or introduce new static analysis warnings.
For example, a naive fix for the running example in Listing~\ref{l:RunningExample} could be to make the field \code{\_number} private and provide getter and setter methods with the names \code{get\_number} and \code{set\_number}.
However, this fix would introduce two new warnings of rule S100: ``Method names should comply with a naming convention'', as the method names do not follow the \code{camelCase} naming convention.

State-of-the-art approaches either do not validate their fixes beyond checking that the targeted warning disappears~\cite{Etemadi2023Sorald:Violations,icse2024-PyTy} or rely on weak validation mechanisms, such as asking an LLM to rank candidate fixes~\cite{Wadhwa2024CORE:LLMs}.
Instead, \name{} addresses this challenge by employing a three-step validation process that (1) builds the project, (2) verifies that the target warning disappears without introducing new warnings, and (3) runs the test suite to check for regressions.

\section{Approach}
\subsection{Overview}

\begin{figure}
	\centering
	\includegraphics[width=1\linewidth]{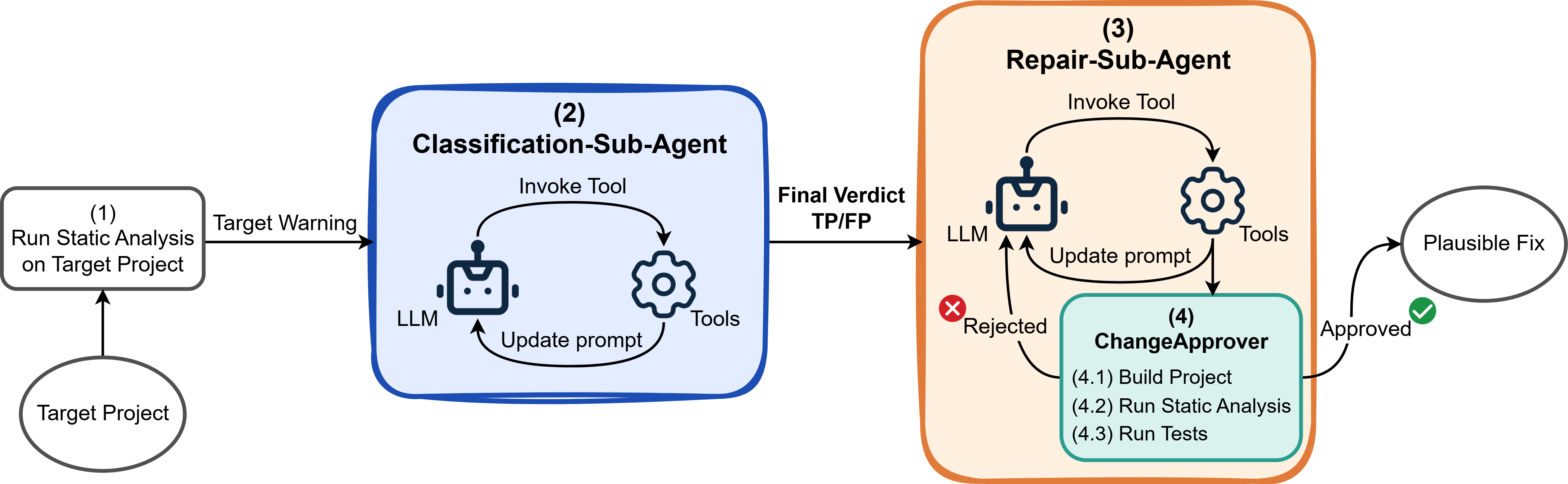}
	\caption{Overview of \name{}.}
	\label{f:overview simple}
\end{figure}

Figure~\ref{f:overview simple} provides an overview of \name{}, which adopts an agentic approach to autonomously classify and fix static analysis warnings.
The approach consists of a preparation stage, followed by two sub-agents: one that classifies a warning as a true positive (TP) or a false positive (FP), and another that fixes or suppresses the warning accordingly.

In the preparation stage (1), \name{} applies static analysis to the target project.
The output is a list of warnings, which \name{} processes one at a time.
For each warning, the \textit{classification sub-agent} (2) determines whether the warning is a TP or FP.
The agent operates in cycles, each consisting of: (i) constructing or updating a dynamic prompt, (ii) querying an LLM with the prompt, and (iii) executing a tool as specified by the LLM response.
Equipped with a toolbox that enables the agent to explore the codebase and the documentation associated with warnings, it issues a classification once sufficient knowledge has been gathered or the computational budget is exhausted.
Once the agent has produced the final verdict, control passes to the second sub-agent.

The \textit{repair sub-agent} (3) edits the codebase to address the warning by either fixing it (in the case of a TP) or suppressing it (in the case of a FP). When the repair sub-agent proposes a patch, the \textit{change approver} (4) validates the plausibility of the fix.
This validation consists of three steps: confirming that the project builds without compilation errors, rerunning the static analyzer to check that the target warning disappears and that no new warnings are introduced, and executing the project's test suite to prevent regressions.
If any check fails, the fix is rejected, and contextual feedback describing the failure is appended to the prompt, enabling the repair sub-agent to gather more information and propose an improved fix.
Once the change approver accepts a fix, it is returned to the user.

The remainder of this section details each of the steps and components of \name{}. 

\subsection{Running Static Analysis on Target Project}
\label{s:preparation steps}

To prepare the input for running \name{}, we run a static analysis on the target project. Static analysis tools may have different output formats and objectives, but in general, they check the program against a set of rules and output a list of locations where these rules are violated.
In our experiments, we use SonarQube as the static analyzer, but \name{} does not rely on any specific property of SonarQube and could be adapted to work with other static analyzers.
We consider a warning to consist of seven fields, as listed in Table~\ref{tab:input_warning_fields}.
The last column of the table provides examples of these fields for the warning in our running example (Listing~\ref{l:RunningExample}).

\begin{table}
	\caption{Input fields given to \name{} with illustrative examples.}
	\centering
	\footnotesize
	\renewcommand{\arraystretch}{1}
	\small
	\begin{tabular}{@{}lp{10em}p{25em}@{}}
		\toprule
		\textbf{Field}          & \textbf{Description}                                 & \textbf{Example} \\ 
		\midrule
		Repository              & Target project at a specific commit                  & https://github.com/uwolfer/gerrit-rest-java-client at de49b4e \\
		RuleKey                 & Key of the violated rule                              & java:S1104 \\
		FilePath                & Path to the file with the warning         & src/main/java/com/google/gerrit/extensions/common/ ChangeInfo.java \\
		StartLine               & Line where the warning is raised                       & 82 \\
		RuleName                & A short description                       & Class variable fields should not have public accessibility \\
		SpecificMessage         & Context-specific warning message                      & Make \scode{\_number} a static final constant or non-public and provide
  accessors if needed. \\
		RuleType                & Type of the rule                                       & Code smell \\
		\bottomrule
	\end{tabular}
	
	\label{tab:input_warning_fields}
\end{table}

\subsection{Classification Sub-Agent: True Positive (TP) vs.\ False Positive (FP)}
Given a target warning, \name{} first determines whether the warning is a TP or a FP.
We place this classification stage before the actual repair because prior work shows that FPs introduce noise and require different handling than TPs~\cite{Kharkar2022,Wen2024,fse2025_suppressions}.
We address this challenge with an agent-based classifier called the \textit{classification sub-agent}.
Consistent with general agent architectures~\cite{icse2025-RepairAgent,Wang2024a}, the classification sub-agent operates in iterative \emph{cycles}, starting by (i) constructing a prompt that includes the task description, the input warning, and the set of accessible tools, (ii) sending the prompt to an LLM, and (iii) parsing the LLM's response into a tool call and executing the tool. These three steps repeat until the classification agent has accumulated sufficient information. At the end, the agent invokes a dedicated tool to emit the final decision on whether the warning is a TP or FP.
Below, we provide further details on prompt design, LLM interactions, and tool invocation.

\subsubsection{LLM Prompting}
Each cycle begins with the explicit construction of a prompt. The prompt is assembled by concatenating the six sections enumerated below, each of which supplies a distinct piece of information to guide the LLM's behavior.

\begin{enumerate}
	\item \textbf{Role and objectives.} This opening segment assigns the agent its role (classifier) and objective: to decide whether the supplied warning is a TP or a FP. By framing the task up front, the model knows exactly what outcome it must produce.
	
	\item \textbf{Constraints.} Here, we introduce constraints and prohibitions. For example, the agent is restricted from seeking user assistance and from introducing new warnings.
	
	\item \textbf{Available tools.} A concise list of the tools the agent may invoke, along with brief usage instructions (Table~\ref{tab:tools_and_their_availability}). Because the LLM must emit its response in a predefined tool-call format, this section acts as a lookup table that maps high-level intents (such as ``read documentation'') to concrete API calls.
	
	\item \textbf{Input warning.} This describes the warning under investigation, as illustrated in Table~\ref{tab:input_warning_fields}. 
	
	\item \textbf{Agent history.} The prompt incorporates a chronologically ordered log of all previous cycles, recording the agent's thoughts, any tool calls made, and the corresponding tool responses. This history serves as the agent's memory, allowing the LLM to build upon earlier deductions rather than starting from scratch in each cycle.
	
	\item \textbf{Definition of response format.} Finally, we define the syntax to be used when issuing a tool call (a JSON-like output containing the tool name and arguments). By enforcing a strict format, downstream processing can reliably parse the output and execute the requested tool.
\end{enumerate}

Given the prompt with these six sections, the LLM is called and its answer is expected to be a tool call in the specified format.

\subsubsection{Tool Invocation}
After the model produces a response, \name{} validates that the response adheres to the defined format. If parsing the response is not possible, the parsing error is appended as feedback to the agent history, helping the agent generate correct responses that can be mapped to a valid tool call in the future. If parsing is successful, the approach invokes the corresponding tool. Once a tool finishes execution, it returns its recorded output. \name{} then adds this output to the agent history, ensuring that subsequent iterations incorporate the result into the prompt. To avoid exceeding the model's context window, the output of a tool invocation is always truncated to the first 125K tokens. This precaution prevents multiple long outputs from exhausting the available context when added to the agent history. An example of tool invocations by the agent over multiple cycles is given in Listing~\ref{fig:trajectory_example_short}. The set of tools enabled for the classification sub-agent is given in Table~\ref{tab:tools_and_their_availability}.

\begin{Listing}
	\centering
	\begin{tcolorbox}[
		colback=darkblue!10!white, colframe=darkblue,
		boxrule=1pt, rounded corners, title=\textbf{Example Agent Trajectory (shortened)},
		boxsep=2pt, top=2pt, bottom=2pt, left=2pt, right=2pt
		]
		\scriptsize
		\textbf{Cycle 1}
		
		\textbf{Tool Call:}  
		ReadDocumentation(rule\_key="S1104")
		
		\textbf{Tool Output:}  
		Class variable fields should not have public accessibility...
		
		---
		
		\textbf{Cycle 2}
		
		\textbf{Tool Call:}  
		ReadLines(file\_path="src/main/.../ChangeInfo.java",  
		start\_line=70, end\_line=90)
		
		\textbf{Tool Output:}  
\begin{lstlisting}[style=JavaCodeStyle, firstnumber=79]
...(shortened)
public int _number;
...(shortened)
\end{lstlisting}
		... (jump to c7)
		
		\textbf{Cycle 7}
		
		\textbf{Tool Call:}  
		AnswerClassificationQuestions(answerQ1="Yes, the rule violation is correctly raised...)
		
		\textbf{Tool Output:}  
		Your answer to question Q1 has been recorded. 

		... (jump to c12)

		\textbf{Cycle 12}

		\textbf{Tool Call:}
		GiveFinalVerdict(verdict="True Positive", rationale="The field \_number is public and accessed directly in multiple files ...")

		\textbf{Tool Output:}
		--- Transition to Repair Sub-Agent ---
	\end{tcolorbox}
	\caption{Shortened agent trajectory with tool calls for the motivating example (Listing~\ref{l:RunningExample}).}
	\label{fig:trajectory_example_short}
\end{Listing}

\begin{table}
	\centering
	\footnotesize
	\caption{Tools available in different parts of \name{}.}
	\renewcommand{\arraystretch}{1}
	\small

	\newcommand{\sectionheader}[1]{%
		\rowcolor{gray!15}%
		\multicolumn{5}{l}{\textbf{#1}} \\[2pt]%
	}
	
	\begin{tabularx}{\textwidth}{
			p{3cm}
			X
			>{\centering\arraybackslash}p{1.cm}
			>{\centering\arraybackslash}p{1.4cm}
			>{\centering\arraybackslash}p{1.7cm}
		}
		\toprule
		& & \multicolumn{3}{c}{\textbf{Sub-Agents}} \\ 
		\cmidrule(l){3-5}
		\textbf{Tool name} & \textbf{Tool description} & \textbf{C-SA} & \textbf{R-SA Fix} & \textbf{R-SA Supp} \\
		\midrule
		
		\sectionheader{Collecting information}
		Read documentation & Retrieves documentation for a given analysis rule. & \checkmark & \checkmark &  \\
		Read lines & Reads a range of lines in a given file. & \checkmark & \checkmark & \checkmark \\ 
		Find references & Finds all project-local references, such as call sites or usages of a symbol. & \checkmark & \checkmark &  \\
		Find definition & Retrieves the definition of a project-local symbol, such as methods and classes. & \checkmark & \checkmark &  \\
		Search patterns & Searches for given patterns. & \checkmark & \checkmark &  \\
		
		\sectionheader{Code Editing}
		Write fix & Used to propose a fix or suppression. &  & \checkmark & \checkmark \\

		\sectionheader{Control tools}
		Formulate plan & Creates/updates warning patching plan. &  & \checkmark &  \\
		Answer classification questions & Gives an answer to the three classification questions. & \checkmark &  &  \\
		Give final verdict & Gives a classification verdict (TP or FP). & \checkmark &  &  \\
		Goals accomplished & Declare end of task with success; allowed only when checks pass.&  & \checkmark & \checkmark \\
		
		\bottomrule
	\end{tabularx}
	
	\vspace{2pt}
	\par\smallskip
	\footnotesize
	\textbf{C-SA} = Classification Sub-Agent \quad\quad
	\textbf{R-SA Fix} = Repair Sub-Agent (Fixing a Warning)\\
	\textbf{R-SA Supp} = Repair Sub-Agent (Suppressing a Warning)

	\label{tab:tools_and_their_availability}
\end{table}

\subsubsection{Classification}
Static analyzers are known to generate large numbers of warnings, many of which are irrelevant or difficult to act upon. This makes classification a crucial first step: distinguishing TPs from FPs prevents wasted repair attempts and avoids introducing incorrect changes~\cite{Kharkar2022,Wen2024}. 

Once the classification agent has collected sufficient information, it can call the tool \textit{Answer classification questions} to address three guiding questions that support the decision on warning classification. In brief, these questions assess: (i) whether the warning is correctly raised and the rule description applies in this specific case, (ii) whether the developer may have intentionally violated the rule (e.g., for functional or design reasons), and (iii) whether the warning can be fixed without breaking existing functionality. The agent gives a yes/no answer and an explanation that refers to collected context to avoid baseless or hallucinated answers.

Following that, the agent issues a final verdict by calling the tool \textit{Give Final Verdict} which asks the agent for the final classification decision and its rationale. Depending on the classification result, \name{} then proceeds to the next stage: if the warning is a TP, the repair sub-agent attempts to fix it, otherwise it attempts to suppress it.

\subsection{Repair Sub-Agent: Fixing or Suppressing the Warning}
The repair sub-agent takes as input the target warning and the final classification decision made by the classification sub-agent. Based on the classification, the sub-agent either fixes the warning or suppresses it. It follows the same agentic approach as the classification sub-agent, but with a different prompt and a tailored set of tools (see Table~\ref{tab:tools_and_their_availability}). 

\subsubsection{Fixing True Positives}
If the warning is classified as a TP, the repair sub-agent attempts to fix it by suggesting modifications to the code. For this purpose, the agent has access to the same information-gathering tools as the classification sub-agent. Using these tools, the agent can retrieve additional code context, examine relevant definitions or references, and identify lines requiring modification. Tools such as \textit{Find references} and \textit{Find definition} are particularly helpful when changes may affect multiple files.  
An additional tool specific to the repair task is \textit{Formulate Plan}. This tool allows the agent to draft a plan for how it intends to fix the warning, which is then added to a dedicated section of the prompt. The plan can be updated iteratively, allowing the agent to adapt to previous attempts' failures.

\begin{Listing}
\begin{lstlisting}[style=jsoncompact]
[{	"file_name": "src/main/java/.../ChangeInfo.java",
	"insertions": [
	{ "line_number": 16,  "new_lines": ["import com.google.gson.annotations.SerializedName;"] },
	{ "line_number": 104, "new_lines": [
		"@SerializedName(\"_number\")",
		"private int number;",
		"public int getNumber() { return number; }",
		"public void setNumber(int n) { this.number = n; }"
		]}
	],
	"deletions": [82]
},
{	"file_name": "src/main/java/.../ChangesRestClient.java",
	"insertions": [
	{ "line_number": 157, "new_lines": ["return id(info.getNumber());"] }
	],
	"deletions": [157]
},
{	"file_name": "src/test/java/.../RealServerTest.java",
	"insertions": [
	{ "line_number": 75, "new_lines": ["int changeNum = changeInfoList.get(0).getNumber();"] },
	{ "line_number": 78, "new_lines": ["Truth.assertThat(change.getNumber()).isEqualTo(changeNum);"] }
	],
	"deletions": [75, 78]
}]
	\end{lstlisting}
	\caption{Shortened fix of motivating example showing insertions, deletions, and multiple files.}
	\label{l:fix_example_short}
\end{Listing}

Once the agent has gathered sufficient information, it calls the \textit{Write fix} tool to propose a modification, expressed in a specified JSON format.
Listing~\ref{l:fix_example_short} shows an example fix for the warning introduced in Listing~\ref{l:RunningExample}. Our running example also illustrates that repairing some warnings may require coordinated changes across multiple files. To do so, insertions and deletions are expressed through the \code{insertions} and \code{deletions} fields for each file.
Once submitted, the proposed fix is applied to the target project and validated using the change approval mechanism described in Section~\ref{s:change approver}.  
If the agent approaches the end of its cycle budget (i.e., five or fewer cycles remaining), a special message is added to the prompt to strongly urge the agent to produce a fix.

\subsubsection{Suppressing False Positives}
If the warning is classified as a FP, the repair sub-agent is tasked with suppressing it. Since this is more straightforward than fixing, the sub-agent is restricted to a smaller toolset: \textit{Read lines}, \textit{Write fix}, and \textit{Goals accomplished}.  
Most static analyzers provide specific mechanisms for suppressing warnings~\cite{fse2025_suppressions}. In our experiments, we use SonarQube, which offers two ways to suppress Java warnings.
The first is to add an inline comment \code{//NOSONAR} to the line where the warning is raised, which suppresses all warnings on that line. The second is to use the \code{@SuppressWarnings(\{"java:S..."\})} annotation on a language construct, such as a method or class, which suppresses all warnings of the specified rule within that scope. We prompt the LLM to prefer the first option, as the second may inadvertently suppress multiple warnings if applied to large scopes, such as classes, and may even suppress warnings that could appear in future code additions~\cite{fse2025_suppressions}. To guide the agent in suppressing warnings, the prompt also includes the explanation given by the classification sub-agent as part of its final verdict.

\subsection{Approving Changes}
\label{s:change approver}

After applying the fix generated by the repair sub-agent, the change approver validates the fix. If the fix is approved, the approach marks it as \emph{plausible} and reports it to the user. Otherwise, the repair sub-agent receives a rejection message with details.  
The change approver evaluates a proposed fix through three validation steps described below.

\subsubsection{Building the Project}
In the first step, the change approver attempts to build the modified project.
If the build fails, the fix is rejected, and the compilation errors are provided as feedback to the agent.
Because build outputs are often lengthy and may contain information from all build substeps, the approach extracts only the relevant compilation errors and ignores standard output. For each error, it reports the file path, line number, the code line that triggered the error, and a diff-like view between the original code and the failing fix.

\subsubsection{Running the Static Analysis}
If the build succeeds, the change approver further validates the fix by running the static analysis on the modified codebase.
This step is inspired by techniques that learn patches from historical fixes to satisfy static analyzer rules~\cite{Bavishi2019,oopsla2019}.
Validation is successful if the target warning has been removed from the analysis report and no new warnings have been introduced.
Otherwise, the fix is rejected.

The location of a warning may change after applying a fix due to inserted or deleted lines.
To accurately determine whether the target warning has been removed and whether new warnings have been introduced, our approach tracks all line changes in modified files and maps warnings between the original and modified lines.
Using this mapping, the approach checks whether the target warning is still present in the corresponding lines after applying the fix. If not, it confirms that the warning has been successfully removed. Similarly, the approach detects newly introduced warnings by mapping warnings from the updated report back to the original lines. If no corresponding entry exists, the approach considers the warning as new.
The approach is sound in the sense that it always detects newly introduced warnings. However, it may over-approximate new warnings in corner cases. For example, if the agent deletes a line with a warning and re-inserts the same line at another location, the warning is flagged as newly introduced.

\subsubsection{Running the Tests}
In the final step, the change approver runs the project's test suite. If all tests pass, the fix is approved and recorded as plausible. If any tests fail, the fix is rejected. In such cases, the change approver extracts the failure reports from the test execution output and cleans them by removing all stack-trace parts outside the target project. The cleaned reports are then provided as feedback to the agent. Using this information, the agent can attempt to refine its fix to avoid regressions, or alternatively adjust failing test cases if they reflect incorrect expectations.

\section{Evaluation}

We evaluate \name{} by addressing the following research questions:

\newlist{researchquestions}{enumerate}{1}
\setlist[researchquestions]{label*=\textbf{RQ\arabic*}}

\begin{researchquestions}    
	\item What is the effectiveness of \name{} at fixing static analysis warnings?
	
	\item How efficient is \name{} with regard to execution time and monetary cost?
	
	\item How does \name{} compare to state-of-the-art approaches?
	
	\item How do different components of the approach contribute to its effectiveness?
	
\end{researchquestions}    

\subsection{Evaluation Setup}

\subsubsection{Dataset}
\label{s:dataset}
We evaluate \name{} on a dataset of SonarQube warnings that we create by running the analysis on 132 open-source Java projects used in the Sorald dataset~\cite{Etemadi2023Sorald:Violations}.
The Sorald dataset consists of 161 projects, from which we exclude four projects to avoid bias, as these were randomly selected to test and improve our approach during development. 
In addition, we filter out 25 other projects that cannot be successfully built with Maven or have failing tests. For the remaining 132 Java projects, we fix the projects to the same commits used by Sorald.
We use the default SonarWay profile for analysis, which contains 479 rules. By running the analysis, we obtain a total of \num{95083} warnings covering 291 distinct rules. 
We aim to evaluate \name{} on a representative sample of these warnings, covering as many distinct rules as possible, to make sure that \name{} can address arbitrary rules, while also reflecting the distribution of warnings in real-world projects. To achieve this, we first randomly sample one warning for each of the 291 distinct rules, ensuring that we cover a wide variety of rules and violation contexts. This results in an initial set of 291 warnings. To further ensure that our evaluation reflects the real-world distribution of warnings, we then randomly sample an additional 709 warnings. By combining these two sampling strategies, we obtain a dataset of \num{1000} warnings spanning 106 projects that we use to evaluate \name{}.

The static analysis classifies rules into four categories: code smell, bug, security hotspot, and vulnerability.\footnote{See \url{https://docs.sonarsource.com/sonarqube-community-build/quality-standards-administration/managing-rules/rules}}
In our dataset, code smells dominate, accounting for 87.7\% of all warnings. The remaining warnings include 9.2\% bugs, 1.9\% security hotspots, and 1.2\% vulnerabilities.
This is close to the distribution in the full set of \num{95083} warnings (95.4\% code smells, 3.0\% bugs, 1.5\% security hotspots, and 0.1\% vulnerabilities).
The security categories (i.e., security hotspots and vulnerabilities) include rules such as ``Delivering code in production with debug features activated is security-sensitive'', which can lead to information leakage, and ``Server hostnames should be verified during SSL/TLS connections'', which, when violated, can make the application susceptible to identity spoofing.
A full list of covered security-related rules, can be found in the \href{https://github.com/sola-st/CodeCureAgent/blob/v0.1.2/security_rules.md}{replication package}.

\subsubsection{LLM and Hardware}
We implement \name{} on top of the AutoGPT framework and RepairAgent~\cite{icse2025-RepairAgent}.
As the LLM, we use GPT-4.1 mini (version 2025-04-14) from OpenAI.
At the time of writing, the cost of the model is \$0.4 per million input tokens and \$1.6 per million output tokens.
The maximum number of cycles is set to 20 and 40 for the classification and repair sub-agents, respectively.
We run all experiments in a Docker container on a virtual machine with an Intel Xeon CPU at 2.1 GHz configured with 16 virtual CPUs and 32 GB of RAM.

\subsubsection{Baselines}

We compare \name{} against three state-of-the-art techniques.

\emph{Sorald}~\cite{Etemadi2023Sorald:Violations} is a rule-based technique that implements fixers for 30 analysis rules.
When a warning matches a supported rule, Sorald applies the corresponding hand-crafted transformation template. This design makes Sorald straightforward and fast on its covered rules, but coverage is inherently limited, and some rules are only partially supported.

\emph{iSMELL}~\cite{Wu2024iSMELL}, an approach focusing on three specific code smell rules, consists of two phases: (1) A detection phase that uses a trained mixture of experts model to select the most accurate static analysis tool for a given code context and code smell rule, to then use this analysis tool to check for the code smell. 
(2) A refactoring phase that, based on a detected code smell and an example of how the code smell rule can be fixed, queries an LLM to generate a fix.
The first phase is orthogonal to our work, and hence, we only compare against the second phase of iSMELL.
We use the publicly available reproduction repository\footnote{See \url{https://github.com/iSMELL2024/iSMELL/tree/0dde27c2264354df80cebe379dda9da8f393d6aa}}, which supports modifying a single file at a time.
Because the refactoring prompts are hard-coded for the three specific code smells targeted by iSMELL, we generalize them to address any SonarQube rule by prompting the LLM with the rule documentation and the full-file warning context.
If the rule documentation does not include an example, we use the LLM to generate one.

Finally, \emph{CORE}~\cite{Wadhwa2024CORE:LLMs} is a recent LLM-based technique that can target arbitrary warnings, provided the corresponding rules and their documentation are supplied in a metadata file.
For each warning, CORE queries an LLM to propose multiple alternative patches that are ranked using an LLM. Their evaluation counts the warning as ``repaired'' if at least one candidate patch removes the warning, unlike our change approver, which executes three checks (build, warning removal, and testing).
A key design characteristic of CORE is that it can modify a single source file only.
CORE was evaluated on a Python dataset with warnings raised by 52 CodeQL static analysis checks and a subset of Sorald's Java dataset with SonarQube, which is limited to 10 different SonarQube rules.
We compare against CORE and the other two baselines on our own Java dataset (Section~\ref{s:dataset}), derived from the set of projects used in Sorald, which covers 291 distinct SonarQube rules.
Comparing against CORE on the Python dataset is beyond the scope of this work, as our implementation of \name{} is specific to Java.
We discuss the necessary adaptations to support other languages and static analyzers in Section~\ref{s:threats}.

For a fair comparison, we run iSMELL and CORE with the same LLM (GPT-4.1 mini) as \name{}.

\subsubsection{Manual Inspection}
\label{s:manual_inspection}
To evaluate both the correctness of the classification and the proposed fixes, we manually inspect the results for the 291 warnings sampled from all violated rules (Section~\ref{s:dataset}).
Due to the high annotation effort, the warnings and fixes are annotated by a single author of the paper, but unclear cases are discussed between the authors.  
During annotation, we follow a structured process:
For the classification, we examine the warning, its context, and the sub-agent's decision-making process to judge whether the reasoning is sound and consistent with the code and the documentation of the static analysis. To inspect the fixes, we review all modified lines to assess whether TP warnings are resolved in accordance with rule documentation or, in the case of FPs, whether the fix correctly suppresses the warning. We also ensure the revised code remains semantically equivalent to the original, unless intentional differences are justified by addressing an underlying bug.
A more detailed description of the annotation process and guidelines is provided in the \href{https://github.com/sola-st/CodeCureAgent/blob/v0.1.2/code_cure_agent/experimental_setups/review_process_definition.md}{replication package}.

\subsection{RQ1: Effectiveness}

\begin{table}[t]
	\centering
	\footnotesize
	\caption{Results of \name{} on the classification and repair task. “Overall” reports results on all \num{1000} warnings; “Manually verified” reports on the 291 inspected warnings. }
	
	\renewcommand{\arraystretch}{1}
	\begin{tabular}{l | rr | rrr}
		\toprule
		& \multicolumn{2}{c|}{\textbf{Overall (/1,000)}} & \multicolumn{3}{c}{\textbf{Manually verified (/291)}} \\
		\cmidrule(lr){2-3} \cmidrule(l){4-6}
		\textbf{Category} 
		&\makecell{\textbf{Agent}\\\textbf{Classification}} 
		& \textbf{Plausible Fix} 
		& \makecell{\textbf{Correct}\\\textbf{Classification}} 
		& \makecell{\textbf{Plausible}\\\textbf{Fix}} 
		& \makecell{\textbf{Correct}\\\textbf{Fix}} \\
		\midrule
		True Positive   & 696 & 665 (95.6\%) & 186/191 (97.4\%) & 178 (93.2\%) & 170 (89.0\%) \\
		False Positive & 304 & 303 (99.7\%) & 81/100 (81.0\%)  & 81 (81.0\%)  & 81 (81.0\%) \\
		\midrule
		\textbf{All warnings} & \textbf{1,000} & \textbf{968 (96.8\%)} & \textbf{267 (91.8\%)} & \textbf{259 (89.0\%)} & \textbf{251 (86.3\%)} \\
		\bottomrule
	\end{tabular}
	
	\label{tab:eval_overall_and_manual}
\end{table}

Table~\ref{tab:eval_overall_and_manual} summarizes the results: Out of \num{1000} warnings, the agent classifies 696 as TP and 304 as FP, and produces plausible fixes for 968 warnings (96.8\%). The plausible-fix rate within each category is similarly high: 665/696 (95.6\%) for TPs and 303/304 (99.7\%) for FPs. This result indicates that once a warning is labeled, the repair sub-agent is highly effective at addressing both TPs and FPs.

Manual inspection of a representative subset of 291 warnings provides a more precise assessment of the agent's performance by checking both the correctness of the classification and the proposed fixes.
We find that the agent correctly classifies 267 of the 291 cases (91.8\%). In the repair step, \name{} attempts fixes for all warnings, but in our manual validation we only judge fixes for correctly classified warnings.
Specifically, \name{} proposes plausible fixes for 259 warnings (89.0\%), of which 251 are confirmed as correct, yielding an overall accuracy of 86.3\%. Most errors arise in the classification phase (8.2\%), rather than in the subsequent repair, which suggests that further improving the classification could positively influence the overall repair success.

Breaking down the manual results by warning class further highlights this pattern.
Among the TPs, the agent classifies 97.4\% correctly, generates plausible fixes for 93.2\%, and correctly fixes 89.0\%.
For FPs, the agent classifies 81.0\% correctly, with the same percentage for plausible fixes and correct fixes.
Thus, the main source of failure is false negatives at the classification stage: \name{} tends toward missed repairs rather than unwanted changes, since suppression does not alter program behavior beyond silencing the warning.

\begin{Listing}
	\begin{tcolorbox}[
		colframe=black,
		colback=white,
		sharp corners,
		boxrule=0.0pt,
		boxsep=0.0pt,
		left=0.2pt,
		right=0.2pt,
		top=5pt,
		bottom=0.0pt
		]
		\scriptsize
		
		\tikz[remember picture]
		\node[anchor=west, baseline, yshift=-0pt] (import_label)
		{Changes to target file \lstinline$MetricsStream.java$};
		
		\begin{tcolorbox}[
			colframe=white,
			colback=white,
			sharp corners,
			boxrule=0.0pt,
			left=0.1pt,
			right=0.1pt,
			top=-10pt,
			bottom=-10pt
			]
			\begin{lstlisting}[style=DiffWithCommentCodeStyle, numbers=none]
void printPerformanceMetrics() {
-    printf( "%s %d\n", getCountName(), scrapeCount);
-    printf(Locale.US, "%s %.2f\n", getDurationName(), toSeconds(getElapsedTime()));
-    printf(Locale.US, "%s %.2f\n", getCpuUsageName(), toSeconds(getCpuUsed()));
+    printf( "%s %d%n", getCountName(), scrapeCount);
+    printf(Locale.US, "%s %.2f%n", getDurationName(), toSeconds(getElapsedTime()));
+    printf(Locale.US, "%s %.2f%n", getCpuUsageName(), toSeconds(getCpuUsed()));
}
			\end{lstlisting}
\end{tcolorbox}
		
\tikz[remember picture]
\node[anchor=north] (settings_label)
{Changes to referencing test file \lstinline$MetricsStreamTest.java$};
		
\begin{tcolorbox}[
			colframe=white,
			colback=white,
			sharp corners,
			boxrule=0.0pt,
			left=0.2pt,
			right=0.2pt,
			top=-10pt,
			bottom=-5pt
			]
\begin{lstlisting}[style=DiffWithCommentCodeStyle, numbers=none]
private List<String> getPrintedMetricValues() {
-    return Arrays.stream(getPrintedMetrics().split("\n")).map(l -> l.split(" ")[1]).collect(Collectors.toList());
+    return Arrays.stream(getPrintedMetrics().split("\\r?\\n")).map(l -> l.split(" ")[1]).collect(Collectors.toList());
}
			\end{lstlisting}
		\end{tcolorbox}
		
	\end{tcolorbox}
	
	\caption{Example of \name{} fix that successfully resolves the true positive warning of rule S3457 in project \href{https://github.com/oracle/weblogic-monitoring-exporter/blob/9d17454fd9e34f05acbf942418c3a92ef40fdec1/wls-exporter-core/src/main/java/com/oracle/wls/exporter/MetricsStream.java\#L72}{weblogic-monitoring-exporter}.}
	\label{l:ExampleFixedWarningFaultyTestFoundAndAdapted}
\end{Listing}

Listing~\ref{l:ExampleFixedWarningFaultyTestFoundAndAdapted} shows an example of a warning that \name{} successfully fixes. 
The warning is raised by rule \code{S3457}: ``Printf-style format strings should be used correctly,'' which points out that \code{\%n} should be used in place of \code{\textbackslash n} to produce a platform-specific line separator. 
\name{} correctly classifies this warning as a TP and then replaces three usages of \code{\textbackslash n} with \code{\%n} within statements in \code{MetricsStream.java}. 
However, after making this change, a test that simulates execution on Windows fails, as it mistakenly uses only \code{\textbackslash n} to split the output line by line, instead of the expected \code{\textbackslash r\textbackslash n} on Windows. 
\name{} inspects the failing test, identifies the issue, and creates a second fix that additionally modifies the test so that it splits the output by arbitrary line separators. 
In this way, \name{} not only resolves the warning but also corrects the faulty test.

\begin{Listing}
	\begin{lstlisting}[style=DiffWithCommentCodeStyle, literate={+0200}{{\color{black}+0200}}{5}, numbers=none, xleftmargin=5pt]
 public void parse(Message message, String line) throws ParseException
 {
 	if( !line.isEmpty() ) {
-   	message.setDate(date_format.parse(line));
+       try {
+       	ZonedDateTime zdt = ZonedDateTime.parse(line, date_format);
+           message.setDate(Date.from(zdt.toInstant()));
+       } catch (DateTimeParseException e) {
+       	throw new ParseException(e.getMessage(), 0);
+       }
	}
+}
 }  // <=== !!Duplicate closing bracket not deleted in the fix
\end{lstlisting}
\caption{Example of \name{} failed fix attempt because of duplicated bracket (\href{https://github.com/lolo101/MsgViewer/blob/42266b9a62ec9706b1042f118fd6f57c6f074fd2/MSGViewer/src/main/java/net/sourceforge/MSGViewer/factory/mbox/headers/DateHeader.java\#L35}{MsgViewer}).}
\label{l:CompilationFailureDueToUndeletedClosingBrace}
\end{Listing}

An example in which the agent fails to identify a correct fix is shown in Listing~\ref{l:CompilationFailureDueToUndeletedClosingBrace}. 
The warning to address is \code{S2885}: ``Non-thread-safe fields should not be static''. 
To fix the warning, the agent replaces the static usage of the non-thread-safe \code{SimpleDateFormat} with the thread-safe \code{DateTimeFormatter}, and \name{} attempts to adapt the local call site accordingly. 
It inserts the full body of the changed method, including an extra closing bracket, thereby duplicating it and breaking the syntactic structure of the class. 
From the failure feedback provided by the change approver, \name{} identifies the issue but still fails to correct it within the allotted cycle budget.
Future work on expressing code changes without breaking the syntactic structure of the code may help to avoid such issues.

\subsubsection{Complexity of Generated Fixes}

\begin{table}[t]
	\centering
	\footnotesize
	\caption{Complexity of fixes and plausible-fix rate for \num{1000} warnings.}
	\renewcommand{\arraystretch}{1}
	\begin{tabular}{l | c c c c}
		\toprule
		\textbf{Fix Complexity} 
		& \textbf{All} 
		& \textbf{TP} 
		& \textbf{FP} 
		& \textbf{Plausible} \\
		\midrule
		\textbf{Single-Line} & 52.0\% & 33.2\% & 95.1\% & 99.4\% \\
		\textbf{Multi-Line}  & 44.4\% & 61.6\% &  4.9\% & 95.5\% \\
		\textbf{Multi-File}  &  3.6\% &  5.2\% &  0.0\% & 75.0\% \\
		\bottomrule
	\end{tabular}
	
	\label{tab:eval_fix_complexity_combined}
\end{table}

To better understand the nature of the fixes generated by \name{}, Table~\ref{tab:eval_fix_complexity_combined} shows the distribution of fixes based on their complexity. Out of \num{1000} fixes, 520 affect only a single line, 444 are multi-line fixes, and 36 are multi-file fixes.
For warnings classified as FPs, almost all fixes are single-line (95.1\%). This is expected, as FPs can usually be suppressed by adding a \code{NOSONAR} comment to the warning line.
In rare cases, however, the suppression spans multiple lines. This can happen, for example, when the reported statement itself is split over several lines and the suppression comment must be added in a way that covers the entire construct.
As shown in the last column of the table, less complex fixes are more likely to be plausible: 99.4\% of all single-line fixes are plausible, compared to 95.5\% for multi-line fixes and an even lower 75.0\% for multi-file fixes.
This observation shows that while \name{} is capable of generating complex fixes, the likelihood of success decreases as complexity increases.

\subsubsection{Types of Fixed Warnings}
When breaking down the results by rule-category (see Section~\ref{s:dataset}), we find that \name{} performs robustly across all categories, generating plausible fixes for 849 code smells (96.8\%), 90 bugs (97.8\%), 17 security hotspots (89.5\%), and all 12 vulnerabilities (100.0\%).

\subsection{RQ2: Efficiency and Costs}

\begin{figure}
	\centering
	\begin{subfigure}[b]{0.49\linewidth}
		\centering
		\includegraphics[width=1\linewidth]{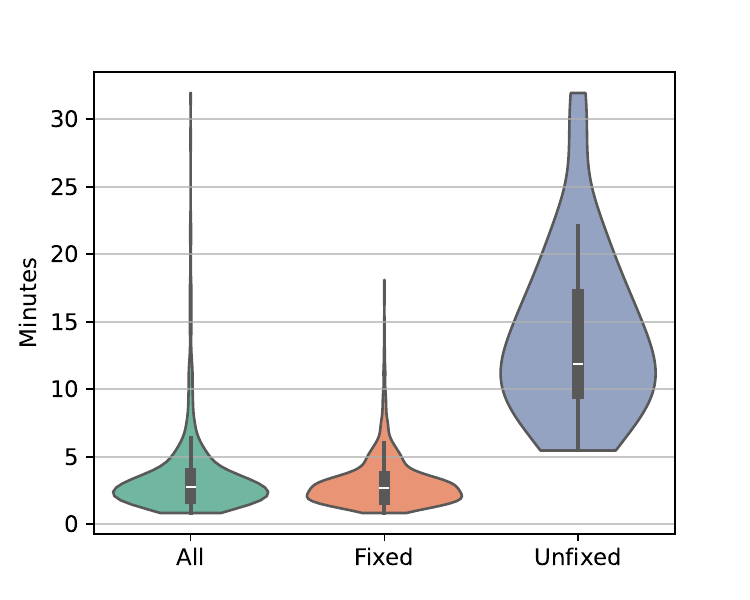}
		\caption{Time}
		\label{f:cca_time_fixed_unfixed}
	\end{subfigure}
	\begin{subfigure}[b]{0.49\linewidth}
		\centering
		\includegraphics[width=1\linewidth]{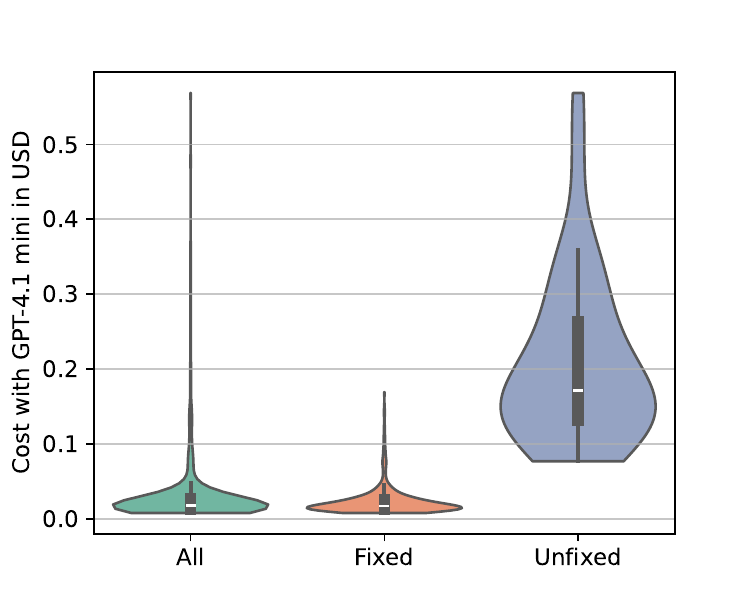}
		\caption{Monetary Cost}
		\label{f:cca_cost_fixed_unfixed}
	\end{subfigure}
	
	\caption{\name{} time and monetary cost distribution between fixed and unfixed warnings.}
	\label{f:cca_time_cost_fixed_unfixed}
\end{figure}

\subsubsection{Time}

Figure~\ref{f:cca_time_fixed_unfixed} shows the distribution of time taken by \name{} for all warnings, fixed, and unfixed ones. On average, \name{} takes 4.4 minutes (median=2.8) to go through a warning. For unfixed warnings, the mean is higher at 14.0 minutes. This is expected, as warnings that the agent fails to fix consume the full repair budget of 40 cycles, compared to an average of only 8 repair cycles for fixed warnings.
A large fraction of \name{}'s execution time is spent outside of running the main agent loop. On average, 41.1\% of the time is attributed to building and testing the target project, and to running the static analysis. In other words, \name{}'s execution time largely depends on the complexity of the project, where projects with complex setups or computationally expensive test suites will need more time to complete.
Another 39.2\% of time is spent querying the LLM. Using faster models could therefore significantly improve the total execution time.

\subsubsection{Token Consumption}
The LLM usage of \name{} incurs a mean token consumption of 139K tokens per warning. Only 3.0\% of these tokens are output tokens (which are usually the more expensive tokens). Of the input tokens, 78.8\% are cached input tokens, since large parts of the prompts created by \name{} are unchanged from previous cycles. At the time of writing, cached input tokens are four times less expensive than uncached input tokens.
This translates to a mean monetary cost for querying the LLM of 2.9 cents (USD) per warning.
Figure~\ref{f:cca_cost_fixed_unfixed} compares the monetary cost between fixed and unfixed warnings. Unfixed warnings have a mean cost of 21 cents. This increased cost aligns with the previous findings that unfixed warnings lead to longer execution time and more agent cycles.

\subsection{RQ3: Comparison to Baselines}
\label{s:comparison_baselines}

\subsubsection{Comparing Effectiveness}
\begin{table}[t]
	\centering
	\footnotesize
	\caption{Unified comparison with baselines. ``Overall (\num{1000})'' reports results relative to the full dataset. ``Sorald-62'' are those 62 warnings supported by Sorald. ``Sorald-Manual-21'' shows the end-to-end correctness on the 21 Sorald-supported warnings that we manually inspected.}
	\renewcommand{\arraystretch}{1}
	\begin{tabular}{l | rr | rr | r}
		\toprule
		& \multicolumn{2}{c|}{\textbf{Overall (1,000)}} 
		& \multicolumn{2}{c|}{\textbf{Sorald-62}} 
		& \makecell{\textbf{Sorald-Manual-21}} \\
		\cmidrule(lr){2-3}\cmidrule(lr){4-5}
		\textbf{Approach}
		& \makecell{\textbf{Created}\\\textbf{Fix}}
		& \makecell{\textbf{Plausible}\\\textbf{Fix}}
		& \makecell{\textbf{Created}\\\textbf{Fix}}
		& \makecell{\textbf{Plausible}\\\textbf{Fix}}
		& \makecell{\textbf{Correct}\\\textbf{Fix}} \\
		\midrule
		\textbf{Sorald}
		& 55 (5.5\%) & 43 (4.3\%)
		& 55 (88.7\%) & 43 (69.4\%)
		& 16 (76.2\%) \\
		\textbf{iSMELL}
		& \num{1000} (100\%) & 628 (62.8\%)
		& 62 (100\%) & 48 (77.4\%) 
		& 13 (61.9\%)\\
		\textbf{CORE}
		& 926 (92.6\%) & 676 (67.6\%)
		& 57 (91.9\%) & 47 (75.8\%)
		& 14 (66.7\%) \\
		\textbf{\name{}}
		& \num{1000} (100\%) & 968 (96.8\%)
		& 62 (100\%) & 62 (100\%)
		& 20 (95.2\%) \\
		\bottomrule
	\end{tabular}
	
	\label{tab:comp_baselines_merged}
\end{table}

Table~\ref{tab:comp_baselines_merged} summarizes the results of comparing \name{} against Sorald, iSMELL, and CORE.
On the full set of \num{1000} warnings, \name{} produces fixes for all warnings and achieves a plausible-fix rate of 96.8\%. 
In contrast, iSMELL drops from 100\% \emph{created} to only 62.8\% \emph{plausible} (a 37.2\% decline), and CORE drops from 92.6\% \emph{created} to 67.6\% \emph{plausible} (a 25.0\% decline), indicating that many candidate fixes fail the oracle checks (builds, tests, or introduction of new warnings). 
Sorald contributes fixes for only 5.5\% of the \num{1000} warnings overall, reflecting its limited coverage of the static analysis rules, with only 4.3\% plausible fixes after validation.

To investigate the performance on the repair task in isolation of classification, we compare the plausible-fix rates on the subset of warnings that \name{} classifies as TPs, which are the warnings that require code changes beyond simple suppression.
\name{} achieves a plausible-fix rate of 95.6\% on TPs, while Sorald, iSMELL, and CORE have 4.6\%, 67.4\%, and 75.6\%, respectively. This shows that even when focusing only on the TPs, \name{} still outperforms the baselines by a significant margin.

The \emph{Sorald-62} part of Table~\ref{tab:comp_baselines_merged} provides a comparison on all 62/\num{1000} warnings that correspond to rules supported by Sorald.
\name{} attains 100\% for both \emph{created} and \emph{plausible} fixes, whereas Sorald reaches 88.7\% created and 69.4\% plausible fixes.
That is, even for the Sorald-supported warnings, \name{} shows a 30.6\% advantage in plausible fixes.

The right-most part of the table shows the results of manually inspecting all 21 warnings from the Sorald-62 set that are part of the manually verified subset of 291 warnings (Section~\ref{s:dataset}).
\name{} correctly repairs 20/21 of these warnings (95.2\%), compared to 16/21 (76.2\%) for Sorald, 13/21 (61.9\%) for iSMELL, and 14/21 (66.7\%) for CORE. 
The single miss for \name{} stems from a misclassification into the FP category, which is again consistent with the earlier observation that mistakes by \name{} predominantly arise during classification.

These results align with the design of the baselines discussed earlier. 
iSMELL and CORE operate on a single file, which prevents them from addressing multi-file changes.
Furthermore, as all baselines do not check for FPs (iSMELL proposes an intelligent selection of analysis tools, but this can only reduce FPs), they are prone to attempt incorrect fixes. For example, \textit{Sorald-62} contains six FPs where attempting to fix the warning according to the rule would break the code, highlighting the importance of reliable classification once again.
Together with the quantitative evidence in Table~\ref{tab:comp_baselines_merged}, these factors explain why \name{} achieves both broader applicability and higher reliability across 291 types of warnings.

\subsubsection{Comparing Efficiency}

We compare \name{}'s efficiency with that of Sorald, iSMELL, and CORE, which we run on the same experimental setup as our approach. Sorald repairs warnings with a mean execution time of 0.5 minutes per warning. This lower execution time is expected, as Sorald is a template-based approach that applies abstract syntax tree transformations, rather than prompting an LLM multiple times. Sorald is even more time-efficient when running on multiple warnings from the same rule at once.
iSMELL has a mean execution time of 1.1 minutes and costs 0.9 cents per warning, which is substantially lower than \name{}'s time and cost (4.4 minutes and 2.9 cents). This is because iSMELL prompts the LLM only once per warning to generate a fix, without any further validation steps.
CORE's mean execution time is 2.3 minutes per warning, about half that of our approach.
At the same time, CORE incurs approximately 4 cents per warning in LLM costs, which is higher than \name{}'s 2.9 cents.
These results are due to CORE prompting an LLM multiple times per run with outputs that contain large sections of code, averaging 255 lines of code per patch.
In contrast, our approach requires only 73 lines on average for specifying edits, of which only 32 lines indicate changed code lines.
The reason why \name{} takes additional time is that it runs multiple validation steps (build, static analysis, and tests) for approving code changes, which we find beneficial for ensuring high-quality fixes (Section~\ref{s:comparison_baselines}).

\subsection{RQ4: Understanding Different Components of \name{}}

\subsubsection{Importance of Checks Done for Change Approval}

\begin{table}[t]
	\centering
	\footnotesize
	\caption{Ablation of the change approver on \num{1000} warnings. ``Checks enforced'' indicates which validations are applied: \emph{Build} (project compiles), \emph{No Warning} (target removed; no new static analysis warnings), and \emph{Tests} (existing suite passes). \emph{Plausible fixes} count patches that would also be accepted by the full change approver; \emph{False accepts} are fixes that slip through the ablated variant but would be rejected when all checks run.}
	\renewcommand{\arraystretch}{1}
	\begin{tabular}{l | c c c | r | r}
		\toprule
		& \multicolumn{3}{c|}{\textbf{Checks enforced}} 
		& \multicolumn{1}{c|}{\textbf{Plausible fix}} 
		& \textbf{Incorrect fix} \\
		\cmidrule(lr){2-4}\cmidrule(lr){5-6}
		\textbf{Variant} 
		& \makecell{\textbf{Build}} 
		& \makecell{\textbf{No Warning}} 
		& \makecell{\textbf{Tests}} 
		& \makecell{\textbf{/\num{1000}} (\%)} 
		& \makecell{\textbf{False accept}} \\
		\midrule
		\textbf{Full Change Approval}          & \cmark & \cmark & \cmark & 968 (96.8\%) & --  \\
		\textbf{w/o Tests}           & \cmark & \cmark & \xmark & 961 (96.1\%) & 9   \\
		\textbf{w/o Static Analysis \& Tests}    & \cmark & \xmark & \xmark & 861 (86.1\%) & 123 \\
		\textbf{No Change Approval}       & \xmark & \xmark & \xmark & 751 (75.1\%) & 249 \\
		\bottomrule
	\end{tabular}
	
	\label{tab:ablation_compact}
\end{table}

Table~\ref{tab:ablation_compact} shows three variants of the change approver based on the activated checks. Removing a component means the reduced pipeline no longer validates that property before accepting a candidate patch. A ``false accept'' is a patch accepted by the reduced checks but rejected by the \emph{full} change approver because it would have failed in build, tests, or in the static analysis check.

As checks are removed, plausibility drops and incorrect fixes slip through. Omitting only tests has a small effect (961 plausible; 9 false accepts), whereas omitting the static analysis check is far more damaging (861 plausible; 123 false accepts). With no checks at all, plausibility falls to 751 (75.1\%), and 249 incorrect fixes would be admitted, which highlights that the checks and the feedback they provide enable the agent to improve 217 otherwise implausible fixes.

Notably, even in the ``No Change Approval'' setting, our approach still exceeds the best baseline's plausible rate on the full dataset: 75.1\% vs 67.6\% for CORE as shown in Table~\ref{tab:comp_baselines_merged}, \emph{despite} CORE being allowed to generate up to 10 candidate patches per warning and counting the fix as “plausible” if \emph{at least one} candidate passes. In our evaluation, \name{} stops at the \emph{first} plausible patch without generating multiple alternatives, yet \name{} demonstrates better plausible fix generation mainly due to dynamic and incremental context collection, focusing on a single warning at a time as opposed to CORE, and benefiting from the classification phase to avoid suggesting code changes for warnings that should simply be suppressed.

\subsubsection{Tools Usage}
\begin{figure}
    \centering
    \includegraphics[width=\linewidth]{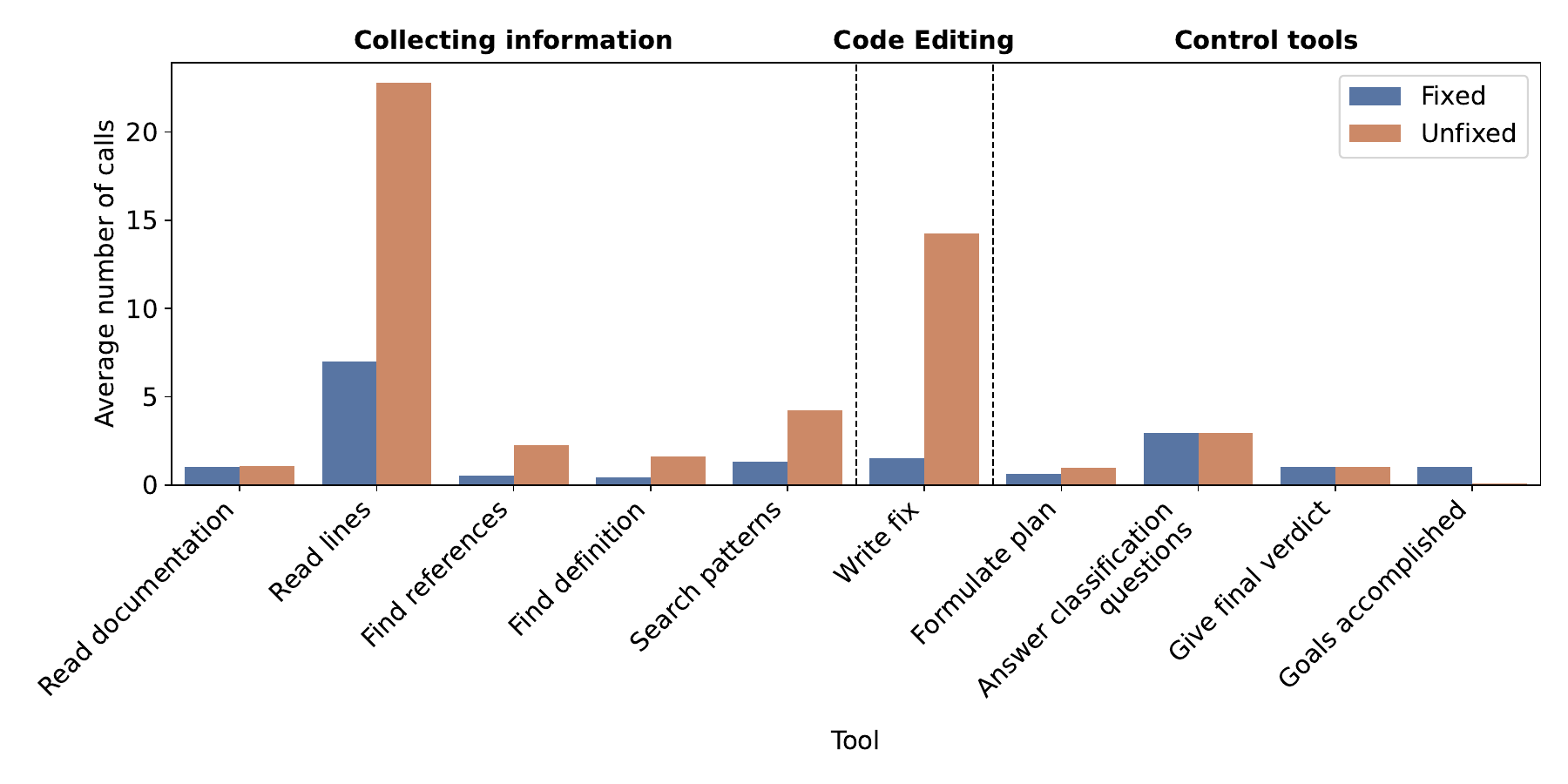}
    \caption{Absolute number of tool calls, comparing between fixed and unfixed warnings.}
    \label{f:average_tool_usage}
\end{figure}
Figure~\ref{f:average_tool_usage} presents the average number of tool calls for successful and failing instances. \name{} invokes all the provided tools at varying rates. Among the information-gathering tools, \textit{Read lines} is used most frequently. The tools \textit{Find references} and \textit{Find definition} are used less often, as they are only needed for certain rules where references or definitions may need to be changed or may influence how the warning is addressed.

The \textit{Answer classification questions} tool is called three times and the \textit{Give final verdict} tool once on average. This is expected, as there are three questions to answer and one final verdict to give in each run of the classification sub-agent.

The \textit{Write fix} tool is called 1.52 times for fixed warnings and 14.25 times for unfixed warnings. This tool is always needed to propose fixes and will be called multiple times if one or more fix attempts are rejected by the change approver. The average number of calls to the \textit{Write fix} tool therefore corresponds to the number of fixes created per warning.

\subsubsection{Causes of success}
The following two examples illustrate how \name{} operates in practice and how its components contribute to successfully resolving warnings.

The first example is a security warning by rule S5042: ``Expanding archive files without controlling resource consumption is security-sensitive'' in \href{https://github.com/opentripplanner/OpenTripPlanner/blob/cc7de54e949771dca9cb7e91b1f0456ae141e9b6/src/main/java/org/opentripplanner/datastore/file/ZipFileDataSource.java\#L82}{OpenTripPlanner}.
\name{} identifies it as a TP, as unchecked expansion can allow for zip bomb attacks. 
The repair sub-agent then tries to create a fix by adding thresholds for the archive file size and compression ratio, but multiple attempted fixes introduce syntax errors. 
The change approver's feedback guides the agent towards improved fixes until the build check passes. 
However, the change approver reports a newly introduced warning, as the fix duplicates a string literal multiple times. The agent reiterates and defines a constant replacing the duplicated literal. 
After passing this change approver step, two tests fail, as they hit the newly added compression ratio threshold. Using the test failure information, the agent increases the threshold accordingly and thereby creates a fix that is accepted by the change approver. 
This example illustrates how \name{} can address challenging warnings by using the change approver checks and feedback to iteratively improve fixes.

\begin{Listing}
\begin{lstlisting}[style=DiffWithCommentCodeStyle, literate={+0200}{{\color{black}+0200}}{5}, numbers=none, xleftmargin=5pt]
public static int compareNewerIsGreater(RecognisedVersion version1, RecognisedVersion version2) {
	return compareNewerIsLower(version2, version1); + //NOSONAR Intentional param inversion
}
\end{lstlisting}
\caption{Example of \name{} suppression of false positive warning of rule S2234 in project \href{https://github.com/toolbox4minecraft/amidst/blob/2902ca61aa8f28e2e3cd903a732b5d29da93b9e3/src/main/java/amidst/mojangapi/minecraftinterface/RecognisedVersion.java\#L342}{Amidst}.}
\label{l:exampleFPInversionOfParameters}
\end{Listing}
In the second example \name{} suppresses a FP, shown in Listing~\ref{l:exampleFPInversionOfParameters}. The warning is about arguments \code{version1} and \code{version2} being passed in reverse order to the \code{compareNewerIsLower} method.
To classify the warning, the classification sub-agent first uses tools to retrieve the rule documentation and the code context around the warning. 
At this point, the agent already suspects that the inversion is intentional, as the method implements the inverted comparison result. To confirm this hypothesis, the agent additionally retrieves all references to \code{compareNewerIsGreater} and \code{compareNewerIsLower}.
Based on this context, the agent is certain that the inversion is intentional, answers the classification questions and gives the final verdict that the warning is a FP, which is then suppressed by the repair sub-agent. 
This example demonstrates how the agent can collect various types of context on demand to make informed decisions and avoid incorrect fixes for FPs.

\subsubsection{Causes of Failures}

\paragraph{\textbf{Causes of wrong classification}}
In the 291 inspected warnings, we find 24 misclassified warnings, which represent 8.2\% of all inspected warnings;
19 wrongly classified as FP.
Incorrect classifications as TP can lead to the fix attempt breaking intended functionality or breaking the build entirely. On the other hand, incorrect  classifications as FP lead to the suppression of an actually valid warning. Upon examination, we find that the failed classifications can be attributed to four main reasons:
\begin{enumerate}
	\item LLM hallucination (11/24): The LLM ignores given context and outputs a decision that is contradictory to indications available in the context.
	
	\item Insufficient context (11/24): Sometimes the static analysis rule is not well documented or imprecise (9/24), and other times the collected context around the warning is shallow (2/24) making the LLM decision almost arbitrary.
	
	\item Other reasons (2/24): Edge cases, where the developer follows a convention that the LLM does not agree with.
\end{enumerate}

\paragraph{\textbf{Causes of inability to repair}}
The repair sub-agent fails to (plausibly) fix 32 warnings.
Furthermore, 8 inspected plausible fixes are incorrect. The failure of repair can be attributed to three main reasons:
(i) Inability to fix incorrect modifications that are breaking code in 26/32 instances (ii) addressing multiple warnings at once in 4/32 instances, despite being instructed to focus on one at a time, and (iii) no support for creating, renaming, or deleting files, meaning a limitation within the available tools in 2/32 cases.

\section{Threats to Validity and Limitations}
\label{s:threats}

We focus on one static analysis (SonarQube) and one programming language (Java).
While both are widely used and highly relevant in practice, the generalizability of our findings to other static analyzers and languages remains to be validated.
To adapt \name{} to another static analyzer, the new analyzer needs to be integrated by making it invokable with a single command, mapping its output to the expected JSON format, and updating the \textit{Read Documentation} tool. 
Adapting \name{} to another programming language requires using a different parser and language server, and to adapt the build and test commands to the new language.
Moreover, code examples in the LLM prompts would need to be updated.
However, the overall approach is not specific to Java and SonarQube, and we expect it to be applicable to other settings.

Our dataset, while diverse, may not capture all types of warnings or project structures, potentially limiting the generalizability of our results.
However, with 291 distinct rules covered, our dataset is an order of magnitude broader than those evaluated by prior work~\cite{Etemadi2023Sorald:Violations,Wadhwa2024CORE:LLMs}, which cover 10 and 52 rules, respectively.

The manual inspection process, despite efforts to ensure consistency, may introduce bias in classification and fix correctness assessments, especially as there is only a single annotator who is one of the authors. To reduce this risk, we follow a systematic annotation process and discuss unclear cases (Section~\ref{s:manual_inspection}).

Additionally, the reliance on GPT-4.1 mini as the LLM may influence results, and different models could yield varying performance.
As stronger models become available and economically viable, we expect \name{}'s performance to improve further.
For a fair comparison with the LLM-based baseline~\cite{Wadhwa2024CORE:LLMs}, we use the same LLM as for \name{}.
Another threat is data leakage. While the LLM might have been pre-exposed to the source code of the projects in our dataset, it has neither seen the exact warnings in the code nor any respective fixes, as the projects generally do not use SonarQube and do not address warnings in dedicated commits. Therefore, the LLM cannot directly recall a fix. Furthermore, the baselines are all evaluated with the same LLM, so any potential data leakage would affect all approaches equally.
Finally, our evaluation primarily focuses on the technical effectiveness, without assessing its impact on developer workflows or acceptance in real-world settings.
Studying how developers interact with and perceive the technique  is an important future work.

\section{Related Work}

\paragraph{\textbf{Static Analysis Tools}}
Range from general-purpose analyzers to domain-specific checkers including SonarQube~\cite{Etemadi2023a}, CodeQL with declarative, multi-language analyses~\cite{Youn2023,issta2023-Fluffy}, and Infer. Some tools are more specific, such as NullAway for Java nullness~\cite{Banerjee2019,Karimipour2023}, test smell detectors~\cite{Wang2021c}, and tensor-shape analysis~\cite{Lagouvardos2020}. Work on learning or synthesizing analysis rules aims to reduce the manual effort of rule creation and maintenance~\cite{Garg2022,Effendi2023}. Complementary efforts build specifications (e.g., taint specs for JS libraries) to strengthen downstream analyses~\cite{icse2020}. Empirical studies further tackle aspects of usage in practice and limitations~\cite{Zheng2021, fse2025_suppressions}.

\paragraph{\textbf{Detecting and Reducing False Positives}}
A first task in mitigation is filtering false alarms. Machine learning techniques classify warnings into TPs and FPs~\cite{Zheng2021,Wen2024}, sometimes validating with targeted tests~\cite{Joshy2021}. Beyond classification, several works reduce analysis imprecision at the source: pruning false call-graph edges~\cite{Utture2022,Reif2019}, neural assistance to refine static analysis~\cite{Zhao2018a,Li2024}. Recent LLM-assisted approaches treat a warning as a structured query containing code and project context~\cite{Wen2024,Li2024}.

\paragraph{\textbf{Automated Repair of Warnings}}
Data-driven repair learns edit patterns from historical fixes and applies them~\cite{Bavishi2019, oopsla2019, Liu2018a}. Rule-based repair (e.g., Sonar quick-fixes and Sorald) shows strong precision but can struggle to generalize beyond encoded idioms~\cite{Etemadi2023a}. StaticFixer augments static warnings with code transformations~\cite{Jain2023}, while synthesis/constraint-driven approaches use analysis goals to guide patching~\cite{Liu2023}. For type-related violations, specialized techniques automatically fix type errors in Python and functional languages~\cite{icse2024-PyTy,Sakkas2020,Oh2022}, and learned type hints can trigger or resolve static checker feedback~\cite{Allamanis2020}. Beyond single-rule scripts, LLM-centric pipelines combine transformation-by-example with static/dynamic checks to produce robust edits~\cite{Dilhara2024}. 

\paragraph{\textbf{General Automated Repair Approaches}} 
A broad line of work in automated program repair has explored how to generate patches for general software defects~\cite{cacm2019-program-repair}. 
Early systems relied on generate and validate strategies such as GenProg~\cite{LeGoues2012} or constraint solving, as in SemFix~\cite{Ke2015}, to search for program variants that satisfy the test suite. 
Template and pattern-based approaches later emerged to capture common fix idioms with higher precision~\cite{Liu2019a,Long2016}, and several such techniques have been deployed in industrial settings~\cite{oopsla2019, marginean}.
The advent of learning-based repair shifted attention toward models trained directly on large corpora of bug fixes~\cite{Tufano2019,Lutellier2020,Chen2021d,Ye2022a,Ye2024} and also type and context-aware repair frameworks~\cite{Jiang2023,Silva2024}. 
More recently, general-purpose large language models have been adapted for repair~\cite{Xia2024a}, and researchers have begun to orchestrate them with agentic prompting pipelines that couple generation with validation and feedback~\cite{icse2025-RepairAgent,Cheng2025}. 
A particularly related work is RepairAgent~\cite{icse2025-RepairAgent}, which also uses an LLM agent with custom designed tools. 
However, RepairAgent targets bug repair with a given bug-revealing test case, while \name{} focuses on fixing static analysis warnings.
Most importantly, RepairAgent assumes that all bugs are TPs, whereas \name{} first classifies warnings, which is crucial for avoiding unnecessary or harmful code changes.
In addition \name{} uses specialized tools for collecting context and validating fixes that are distinct from RepairAgent. Lastly, \name{} is even capable of fixing wrong or inaccurate test cases (e.g., see Listing~\ref{l:ExampleFixedWarningFaultyTestFoundAndAdapted}).

\section{Conclusion}
In this paper, we present \name{}, an autonomous LLM agent that first classifies static analysis warnings as TPs or FPs, then either repairs or suppresses them. A built-in change approver validates each edit by ensuring the project builds successfully, the original warning is eliminated, no new warnings arise, and all tests pass. Evaluated on \num{1000} SonarQube warnings across 106 Java projects, \name{} generates plausible fixes for 96.8\% of warnings, correctly classifies 91.8\% of distinct-rule cases, and achieves an overall end-to-end correct fix rate of 86.3\%, surpassing prior techniques such as Sorald, iSMELL, and CORE. 
By addressing multi-line and multi-file fixes, and by filtering FPs, \name{} significantly reduces the manual effort needed to maintain clean codebases for a cost of only a few cents and a couple of minutes per warning. Future work could expand language support, target new static analyzers, and incorporate developer feedback for continual learning and integration of developer intent.

\section{Data Availability}
The code and data are publicly available at \href{https://github.com/sola-st/CodeCureAgent}{https://github.com/sola-st/CodeCureAgent} and archived with a DOI on Zenodo~\cite{CodeCureAgentZenodo}.

\bibliographystyle{ACM-Reference-Format}
\bibliography{codecurerefs,referencesMichael}


\begin{thebibliography}{59}


\ifx \showCODEN    \undefined \def \showCODEN     #1{\unskip}     \fi
\ifx \showISBNx    \undefined \def \showISBNx     #1{\unskip}     \fi
\ifx \showISBNxiii \undefined \def \showISBNxiii  #1{\unskip}     \fi
\ifx \showISSN     \undefined \def \showISSN      #1{\unskip}     \fi
\ifx \showLCCN     \undefined \def \showLCCN      #1{\unskip}     \fi
\ifx \shownote     \undefined \def \shownote      #1{#1}          \fi
\ifx \showarticletitle \undefined \def \showarticletitle #1{#1}   \fi
\ifx \showURL      \undefined \def \showURL       {\relax}        \fi
\providecommand\bibfield[2]{#2}
\providecommand\bibinfo[2]{#2}
\providecommand\natexlab[1]{#1}
\providecommand\showeprint[2][]{arXiv:#2}

\bibitem[Allamanis et~al\mbox{.}(2020)]%
        {Allamanis2020}
\bibfield{author}{\bibinfo{person}{Miltiadis Allamanis}, \bibinfo{person}{Earl~T. Barr}, \bibinfo{person}{Soline Ducousso}, {and} \bibinfo{person}{Zheng Gao}.} \bibinfo{year}{2020}\natexlab{}.
\newblock \showarticletitle{Typilus: neural type hints}. In \bibinfo{booktitle}{\emph{Proceedings of the 41st {ACM} {SIGPLAN} International Conference on Programming Language Design and Implementation, {PLDI}}}. \bibinfo{pages}{91--105}.
\newblock
\href{https://doi.org/10.1145/3385412.3385997}{doi:\nolinkurl{10.1145/3385412.3385997}}


\bibitem[Bader et~al\mbox{.}(2019)]%
        {oopsla2019}
\bibfield{author}{\bibinfo{person}{Johannes Bader}, \bibinfo{person}{Andrew Scott}, \bibinfo{person}{Michael Pradel}, {and} \bibinfo{person}{Satish Chandra}.} \bibinfo{year}{2019}\natexlab{}.
\newblock \showarticletitle{Getafix: {L}earning to fix bugs automatically}.
\newblock \bibinfo{journal}{\emph{Proc. {ACM} Program. Lang.}} \bibinfo{volume}{3}, \bibinfo{number}{{OOPSLA}} (\bibinfo{year}{2019}), \bibinfo{pages}{159:1--159:27}.
\newblock
\href{https://doi.org/10.1145/3360585}{doi:\nolinkurl{10.1145/3360585}}


\bibitem[Banerjee et~al\mbox{.}(2019)]%
        {Banerjee2019}
\bibfield{author}{\bibinfo{person}{Subarno Banerjee}, \bibinfo{person}{Lazaro Clapp}, {and} \bibinfo{person}{Manu Sridharan}.} \bibinfo{year}{2019}\natexlab{}.
\newblock \showarticletitle{NullAway: practical type-based null safety for Java}. In \bibinfo{booktitle}{\emph{Proceedings of the {ACM} Joint Meeting on European Software Engineering Conference and Symposium on the Foundations of Software Engineering, {ESEC/SIGSOFT} {FSE} 2019, Tallinn, Estonia, August 26-30, 2019}}, \bibfield{editor}{\bibinfo{person}{Marlon Dumas}, \bibinfo{person}{Dietmar Pfahl}, \bibinfo{person}{Sven Apel}, {and} \bibinfo{person}{Alessandra Russo}} (Eds.). \bibinfo{publisher}{{ACM}}, \bibinfo{pages}{740--750}.
\newblock
\href{https://doi.org/10.1145/3338906.3338919}{doi:\nolinkurl{10.1145/3338906.3338919}}


\bibitem[Barei{\ss} et~al\mbox{.}(2022)]%
        {codexStudy2022}
\bibfield{author}{\bibinfo{person}{Patrick Barei{\ss}}, \bibinfo{person}{Beatriz Souza}, \bibinfo{person}{Marcelo d'Amorim}, {and} \bibinfo{person}{Michael Pradel}.} \bibinfo{year}{2022}\natexlab{}.
\newblock \showarticletitle{Code Generation Tools (Almost) for Free? {A} Study of Few-Shot, Pre-Trained Language Models on Code}.
\newblock \bibinfo{journal}{\emph{CoRR}}  \bibinfo{volume}{abs/2206.01335} (\bibinfo{year}{2022}).
\newblock
\showeprint[arXiv]{2206.01335}
\href{https://doi.org/10.48550/arXiv.2206.01335}{doi:\nolinkurl{10.48550/arXiv.2206.01335}}


\bibitem[Bavishi et~al\mbox{.}(2019)]%
        {Bavishi2019}
\bibfield{author}{\bibinfo{person}{Rohan Bavishi}, \bibinfo{person}{Hiroaki Yoshida}, {and} \bibinfo{person}{Mukul~R. Prasad}.} \bibinfo{year}{2019}\natexlab{}.
\newblock \showarticletitle{Phoenix: automated data-driven synthesis of repairs for static analysis violations}. In \bibinfo{booktitle}{\emph{ESEC/FSE}}. \bibinfo{pages}{613--624}.
\newblock
\href{https://doi.org/10.1145/3338906.3338952}{doi:\nolinkurl{10.1145/3338906.3338952}}


\bibitem[Bouzenia et~al\mbox{.}(2025)]%
        {icse2025-RepairAgent}
\bibfield{author}{\bibinfo{person}{Islem Bouzenia}, \bibinfo{person}{Premkumar Devanbu}, {and} \bibinfo{person}{Michael Pradel}.} \bibinfo{year}{2025}\natexlab{}.
\newblock \showarticletitle{{RepairAgent}: An Autonomous, {LLM}-Based Agent for Program Repair}. In \bibinfo{booktitle}{\emph{International Conference on Software Engineering (ICSE)}}.
\newblock


\bibitem[Chen et~al\mbox{.}(2023)]%
        {Chen2023a}
\bibfield{author}{\bibinfo{person}{Xinyun Chen}, \bibinfo{person}{Maxwell Lin}, \bibinfo{person}{Nathanael Schärli}, {and} \bibinfo{person}{Denny Zhou}.} \bibinfo{year}{2023}\natexlab{}.
\newblock \bibinfo{title}{Teaching Large Language Models to Self-Debug}.
\newblock
\showeprint[arxiv]{2304.05128}~[cs.CL]


\bibitem[Chen et~al\mbox{.}(2021)]%
        {Chen2021d}
\bibfield{author}{\bibinfo{person}{Zimin Chen}, \bibinfo{person}{Steve Kommrusch}, \bibinfo{person}{Michele Tufano}, \bibinfo{person}{Louis{-}No{\"{e}}l Pouchet}, \bibinfo{person}{Denys Poshyvanyk}, {and} \bibinfo{person}{Martin Monperrus}.} \bibinfo{year}{2021}\natexlab{}.
\newblock \showarticletitle{{SequenceR}: Sequence-to-Sequence Learning for End-to-End Program Repair}.
\newblock \bibinfo{journal}{\emph{{IEEE} Trans. Software Eng.}} \bibinfo{volume}{47}, \bibinfo{number}{9} (\bibinfo{year}{2021}), \bibinfo{pages}{1943--1959}.
\newblock
\href{https://doi.org/10.1109/TSE.2019.2940179}{doi:\nolinkurl{10.1109/TSE.2019.2940179}}


\bibitem[Cheng et~al\mbox{.}(2025)]%
        {Cheng2025}
\bibfield{author}{\bibinfo{person}{Runxiang Cheng}, \bibinfo{person}{Michele Tufano}, \bibinfo{person}{J{\"u}rgen Cito}, \bibinfo{person}{Jos{\'e} Cambronero}, \bibinfo{person}{Pat Rondon}, \bibinfo{person}{Renyao Wei}, \bibinfo{person}{Aaron Sun}, {and} \bibinfo{person}{Satish Chandra}.} \bibinfo{year}{2025}\natexlab{}.
\newblock \showarticletitle{Agentic Bug Reproduction for Effective Automated Program Repair at Google}.
\newblock \bibinfo{journal}{\emph{arXiv preprint arXiv:2502.01821}} (\bibinfo{year}{2025}).
\newblock


\bibitem[Chow et~al\mbox{.}(2024a)]%
        {Chow_2024}
\bibfield{author}{\bibinfo{person}{Yiu~Wai Chow}, \bibinfo{person}{Luca Di~Grazia}, {and} \bibinfo{person}{Michael Pradel}.} \bibinfo{year}{2024}\natexlab{a}.
\newblock \showarticletitle{PyTy: Repairing Static Type Errors in Python}. In \bibinfo{booktitle}{\emph{Proceedings of the IEEE/ACM 46th International Conference on Software Engineering}} \emph{(\bibinfo{series}{ICSE ’24})}. \bibinfo{publisher}{ACM}, \bibinfo{pages}{1–13}.
\newblock
\href{https://doi.org/10.1145/3597503.3639184}{doi:\nolinkurl{10.1145/3597503.3639184}}


\bibitem[Chow et~al\mbox{.}(2024b)]%
        {icse2024-PyTy}
\bibfield{author}{\bibinfo{person}{Yiu~Wai Chow}, \bibinfo{person}{Luca~Di Grazia}, {and} \bibinfo{person}{Michael Pradel}.} \bibinfo{year}{2024}\natexlab{b}.
\newblock \showarticletitle{PyTy: Repairing Static Type Errors in Python}. In \bibinfo{booktitle}{\emph{Proceedings of the 46th {IEEE/ACM} International Conference on Software Engineering, {ICSE} 2024, Lisbon, Portugal, April 14-20, 2024}}. \bibinfo{publisher}{{ACM}}, \bibinfo{pages}{87:1--87:13}.
\newblock
\href{https://doi.org/10.1145/3597503.3639184}{doi:\nolinkurl{10.1145/3597503.3639184}}


\bibitem[Chow et~al\mbox{.}(2023)]%
        {issta2023-Fluffy}
\bibfield{author}{\bibinfo{person}{Yiu~Wai Chow}, \bibinfo{person}{Max Sch{\"{a}}fer}, {and} \bibinfo{person}{Michael Pradel}.} \bibinfo{year}{2023}\natexlab{}.
\newblock \showarticletitle{Beware of the Unexpected: Bimodal Taint Analysis}. In \bibinfo{booktitle}{\emph{Proceedings of the 32nd {ACM} {SIGSOFT} International Symposium on Software Testing and Analysis, {ISSTA}}}, \bibfield{editor}{\bibinfo{person}{Ren{\'{e}} Just} {and} \bibinfo{person}{Gordon Fraser}} (Eds.). \bibinfo{publisher}{{ACM}}, \bibinfo{pages}{211--222}.
\newblock
\href{https://doi.org/10.1145/3597926.3598050}{doi:\nolinkurl{10.1145/3597926.3598050}}


\bibitem[Dilhara et~al\mbox{.}(2024)]%
        {Dilhara2024}
\bibfield{author}{\bibinfo{person}{Malinda Dilhara}, \bibinfo{person}{Abhiram Bellur}, \bibinfo{person}{Timofey Bryksin}, {and} \bibinfo{person}{Danny Dig}.} \bibinfo{year}{2024}\natexlab{}.
\newblock \showarticletitle{Unprecedented Code Change Automation: The Fusion of {LLMs} and Transformation by Example}. In \bibinfo{booktitle}{\emph{FSE}}.
\newblock
\urldef\tempurl%
\url{https://doi.org/10.48550/arXiv.2402.07138}
\showURL{%
\tempurl}


\bibitem[Effendi et~al\mbox{.}(2023)]%
        {Effendi2023}
\bibfield{author}{\bibinfo{person}{Sedick David~Baker Effendi}, \bibinfo{person}{Berk Cirisci}, \bibinfo{person}{Rajdeep Mukherjee}, \bibinfo{person}{Hoan Nguyen}, {and} \bibinfo{person}{Omer Tripp}.} \bibinfo{year}{2023}\natexlab{}.
\newblock \showarticletitle{A language-agnostic framework for mining static analysis rules from code changes}. In \bibinfo{booktitle}{\emph{ICSE-SEIP}}.
\newblock


\bibitem[Etemadi et~al\mbox{.}(2023a)]%
        {Etemadi2023a}
\bibfield{author}{\bibinfo{person}{Khashayar Etemadi}, \bibinfo{person}{Nicolas Harrand}, \bibinfo{person}{Simon Lars{\'{e}}n}, \bibinfo{person}{Haris Adzemovic}, \bibinfo{person}{Henry~Luong Phu}, \bibinfo{person}{Ashutosh Verma}, \bibinfo{person}{Fernanda Madeiral}, \bibinfo{person}{Douglas Wikstr{\"{o}}m}, {and} \bibinfo{person}{Martin Monperrus}.} \bibinfo{year}{2023}\natexlab{a}.
\newblock \showarticletitle{Sorald: Automatic Patch Suggestions for SonarQube Static Analysis Violations}.
\newblock \bibinfo{journal}{\emph{{IEEE} Trans. Dependable Secur. Comput.}} \bibinfo{volume}{20}, \bibinfo{number}{4} (\bibinfo{year}{2023}), \bibinfo{pages}{2794--2810}.
\newblock
\href{https://doi.org/10.1109/TDSC.2022.3167316}{doi:\nolinkurl{10.1109/TDSC.2022.3167316}}


\bibitem[Etemadi et~al\mbox{.}(2023b)]%
        {Etemadi2023Sorald:Violations}
\bibfield{author}{\bibinfo{person}{Khashayar Etemadi}, \bibinfo{person}{Nicolas Harrand}, \bibinfo{person}{Simon Lars{\'{e}}n}, \bibinfo{person}{Haris Adzemovic}, \bibinfo{person}{Henry~Luong Phu}, \bibinfo{person}{Ashutosh Verma}, \bibinfo{person}{Fernanda Madeiral}, \bibinfo{person}{Douglas Wikstr{\"{o}}m}, {and} \bibinfo{person}{Martin Monperrus}.} \bibinfo{year}{2023}\natexlab{b}.
\newblock \showarticletitle{{Sorald: Automatic Patch Suggestions for SonarQube Static Analysis Violations}}.
\newblock \bibinfo{journal}{\emph{IEEE Transactions on Dependable and Secure Computing}} \bibinfo{volume}{20}, \bibinfo{number}{4} (\bibinfo{date}{7} \bibinfo{year}{2023}), \bibinfo{pages}{2794--2810}.
\newblock
\showISSN{1545-5971}
\href{https://doi.org/10.1109/TDSC.2022.3167316}{doi:\nolinkurl{10.1109/TDSC.2022.3167316}}


\bibitem[Garg and Sengamedu(2022)]%
        {Garg2022}
\bibfield{author}{\bibinfo{person}{Pranav Garg} {and} \bibinfo{person}{Srinivasan Sengamedu}.} \bibinfo{year}{2022}\natexlab{}.
\newblock \showarticletitle{Example-based Synthesis of Static Analysis Rules}.
\newblock \bibinfo{journal}{\emph{CoRR}}  \bibinfo{volume}{abs/2204.08643} (\bibinfo{year}{2022}).
\newblock
\showeprint[arXiv]{2204.08643}
\href{https://doi.org/10.48550/arXiv.2204.08643}{doi:\nolinkurl{10.48550/arXiv.2204.08643}}


\bibitem[Gu et~al\mbox{.}(2024)]%
        {Gu2024}
\bibfield{author}{\bibinfo{person}{Alex Gu}, \bibinfo{person}{Wen-Ding Li}, \bibinfo{person}{Naman Jain}, \bibinfo{person}{Theo~X. Olausson}, \bibinfo{person}{Celine Lee}, \bibinfo{person}{Koushik Sen}, {and} \bibinfo{person}{Armando Solar-Lezama}.} \bibinfo{year}{2024}\natexlab{}.
\newblock \bibinfo{title}{The Counterfeit Conundrum: Can Code Language Models Grasp the Nuances of Their Incorrect Generations?}
\newblock
\showeprint[arxiv]{2402.19475}~[cs.SE]


\bibitem[Hu et~al\mbox{.}(2025)]%
        {fse2025_suppressions}
\bibfield{author}{\bibinfo{person}{Huimin Hu}, \bibinfo{person}{Yingying Wang}, \bibinfo{person}{Julia Rubin}, {and} \bibinfo{person}{Michael Pradel}.} \bibinfo{year}{2025}\natexlab{}.
\newblock \showarticletitle{An Empirical Study of Suppressed Static Analysis Warnings}.
\newblock \bibinfo{journal}{\emph{Proceedings of the ACM on Software Engineering}} \bibinfo{volume}{2}, \bibinfo{number}{FSE} (\bibinfo{year}{2025}), \bibinfo{pages}{290--311}.
\newblock


\bibitem[Imtiaz et~al\mbox{.}(2019)]%
        {imtiaz2019challenges}
\bibfield{author}{\bibinfo{person}{Nasif Imtiaz}, \bibinfo{person}{Akond Rahman}, \bibinfo{person}{Effat Farhana}, {and} \bibinfo{person}{Laurie Williams}.} \bibinfo{year}{2019}\natexlab{}.
\newblock \showarticletitle{Challenges with responding to static analysis tool alerts}. In \bibinfo{booktitle}{\emph{2019 IEEE/ACM 16th International Conference on Mining Software Repositories (MSR)}}. IEEE, \bibinfo{pages}{245--249}.
\newblock
\href{https://doi.org/10.1109/MSR.2019.00049}{doi:\nolinkurl{10.1109/MSR.2019.00049}}


\bibitem[Jain et~al\mbox{.}(2023)]%
        {Jain2023}
\bibfield{author}{\bibinfo{person}{Naman Jain}, \bibinfo{person}{Shubham Gandhi}, \bibinfo{person}{Atharv Sonwane}, \bibinfo{person}{Aditya Kanade}, \bibinfo{person}{Nagarajan Natarajan}, \bibinfo{person}{Suresh Parthasarathy}, \bibinfo{person}{Sriram Rajamani}, {and} \bibinfo{person}{Rahul Sharma}.} \bibinfo{year}{2023}\natexlab{}.
\newblock \bibinfo{title}{StaticFixer: From Static Analysis to Static Repair}.
\newblock
\showeprint[arxiv]{2307.12465}~[cs.SE]


\bibitem[Jiang et~al\mbox{.}(2023)]%
        {Jiang2023}
\bibfield{author}{\bibinfo{person}{Nan Jiang}, \bibinfo{person}{Kevin Liu}, \bibinfo{person}{Thibaud Lutellier}, {and} \bibinfo{person}{Lin Tan}.} \bibinfo{year}{2023}\natexlab{}.
\newblock \showarticletitle{Impact of Code Language Models on Automated Program Repair}. In \bibinfo{booktitle}{\emph{ICSE}}. \bibinfo{pages}{1430--1442}.
\newblock
\href{https://doi.org/10.1109/ICSE48619.2023.00125}{doi:\nolinkurl{10.1109/ICSE48619.2023.00125}}


\bibitem[Johnson et~al\mbox{.}(2013)]%
        {johnson2013don}
\bibfield{author}{\bibinfo{person}{Brittany Johnson}, \bibinfo{person}{Yoonki Song}, \bibinfo{person}{Emerson Murphy-Hill}, {and} \bibinfo{person}{Robert Bowdidge}.} \bibinfo{year}{2013}\natexlab{}.
\newblock \showarticletitle{Why don't software developers use static analysis tools to find bugs?}. In \bibinfo{booktitle}{\emph{2013 35th International Conference on Software Engineering (ICSE)}}. IEEE, \bibinfo{pages}{672--681}.
\newblock
\href{https://doi.org/10.1109/ICSE.2013.6606613}{doi:\nolinkurl{10.1109/ICSE.2013.6606613}}


\bibitem[Joos et~al\mbox{.}(2026)]%
        {CodeCureAgentZenodo}
\bibfield{author}{\bibinfo{person}{Pascal Joos}, \bibinfo{person}{Islem Bouzenia}, {and} \bibinfo{person}{Michael Pradel}.} \bibinfo{year}{2026}\natexlab{}.
\newblock \bibinfo{title}{CodeCureAgent: Automatic Classification and Repair of Static Analysis Warnings}.
\newblock \bibinfo{howpublished}{Software and data archived in Zenodo}.
\newblock
\href{https://doi.org/10.5281/zenodo.19552529}{doi:\nolinkurl{10.5281/zenodo.19552529}}


\bibitem[Joshy et~al\mbox{.}(2021)]%
        {Joshy2021}
\bibfield{author}{\bibinfo{person}{Ashwin~Kallingal Joshy}, \bibinfo{person}{Xueyuan Chen}, \bibinfo{person}{Benjamin Steenhoek}, {and} \bibinfo{person}{Wei Le}.} \bibinfo{year}{2021}\natexlab{}.
\newblock \showarticletitle{Validating Static Warnings via Testing Code Fragments}. In \bibinfo{booktitle}{\emph{ISSTA}}.
\newblock


\bibitem[Karimipour et~al\mbox{.}(2023)]%
        {Karimipour2023}
\bibfield{author}{\bibinfo{person}{Nima Karimipour}, \bibinfo{person}{Justin Pham}, \bibinfo{person}{Lazaro Clapp}, {and} \bibinfo{person}{Manu Sridharan}.} \bibinfo{year}{2023}\natexlab{}.
\newblock \showarticletitle{Practical Inference of Nullability Types}. In \bibinfo{booktitle}{\emph{Proceedings of the 31st {ACM} Joint European Software Engineering Conference and Symposium on the Foundations of Software Engineering, {ESEC/FSE} 2023, San Francisco, CA, USA, December 3-9, 2023}}, \bibfield{editor}{\bibinfo{person}{Satish Chandra}, \bibinfo{person}{Kelly Blincoe}, {and} \bibinfo{person}{Paolo Tonella}} (Eds.). \bibinfo{publisher}{{ACM}}, \bibinfo{pages}{1395--1406}.
\newblock
\href{https://doi.org/10.1145/3611643.3616326}{doi:\nolinkurl{10.1145/3611643.3616326}}


\bibitem[Ke et~al\mbox{.}(2015)]%
        {Ke2015}
\bibfield{author}{\bibinfo{person}{Yalin Ke}, \bibinfo{person}{Kathryn~T Stolee}, \bibinfo{person}{Claire Le~Goues}, {and} \bibinfo{person}{Yuriy Brun}.} \bibinfo{year}{2015}\natexlab{}.
\newblock \showarticletitle{Repairing programs with semantic code search (t)}. In \bibinfo{booktitle}{\emph{ASE}}. IEEE, \bibinfo{pages}{295--306}.
\newblock


\bibitem[Kharkar et~al\mbox{.}(2022)]%
        {Kharkar2022}
\bibfield{author}{\bibinfo{person}{Anant Kharkar}, \bibinfo{person}{Roshanak~Zilouchian Moghaddam}, \bibinfo{person}{Matthew Jin}, \bibinfo{person}{Xiaoyu Liu}, \bibinfo{person}{Xin Shi}, \bibinfo{person}{Colin~B. Clement}, {and} \bibinfo{person}{Neel Sundaresan}.} \bibinfo{year}{2022}\natexlab{}.
\newblock \showarticletitle{Learning to Reduce False Positives in Analytic Bug Detectors}. In \bibinfo{booktitle}{\emph{44th {IEEE/ACM} 44th International Conference on Software Engineering, {ICSE}}}. \bibinfo{pages}{1307--1316}.
\newblock
\href{https://doi.org/10.1145/3510003.3510153}{doi:\nolinkurl{10.1145/3510003.3510153}}


\bibitem[Lagouvardos et~al\mbox{.}(2020)]%
        {Lagouvardos2020}
\bibfield{author}{\bibinfo{person}{Sifis Lagouvardos}, \bibinfo{person}{Julian Dolby}, \bibinfo{person}{Neville Grech}, \bibinfo{person}{Anastasios Antoniadis}, {and} \bibinfo{person}{Yannis Smaragdakis}.} \bibinfo{year}{2020}\natexlab{}.
\newblock \showarticletitle{Static Analysis of Shape in TensorFlow Programs}. In \bibinfo{booktitle}{\emph{34th European Conference on Object-Oriented Programming, {ECOOP}}}, Vol.~\bibinfo{volume}{166}. \bibinfo{pages}{15:1--15:29}.
\newblock
\href{https://doi.org/10.4230/LIPIcs.ECOOP.2020.15}{doi:\nolinkurl{10.4230/LIPIcs.ECOOP.2020.15}}


\bibitem[Le et~al\mbox{.}(2022)]%
        {Le2022a}
\bibfield{author}{\bibinfo{person}{Quang~Loc Le}, \bibinfo{person}{Azalea Raad}, \bibinfo{person}{Jules Villard}, \bibinfo{person}{Josh Berdine}, \bibinfo{person}{Derek Dreyer}, {and} \bibinfo{person}{Peter~W. O'Hearn}.} \bibinfo{year}{2022}\natexlab{}.
\newblock \showarticletitle{Finding real bugs in big programs with incorrectness logic}.
\newblock \bibinfo{journal}{\emph{Proc. {ACM} Program. Lang.}} \bibinfo{volume}{6}, \bibinfo{number}{{OOPSLA}} (\bibinfo{year}{2022}), \bibinfo{pages}{1--27}.
\newblock
\href{https://doi.org/10.1145/3527325}{doi:\nolinkurl{10.1145/3527325}}


\bibitem[{Le Goues} et~al\mbox{.}(2012)]%
        {LeGoues2012}
\bibfield{author}{\bibinfo{person}{Claire {Le Goues}}, \bibinfo{person}{ThanhVu Nguyen}, \bibinfo{person}{Stephanie Forrest}, {and} \bibinfo{person}{Westley Weimer}.} \bibinfo{year}{2012}\natexlab{}.
\newblock \showarticletitle{GenProg: {A} Generic Method for Automatic Software Repair}.
\newblock \bibinfo{journal}{\emph{{IEEE} Trans. Software Eng.}} \bibinfo{volume}{38}, \bibinfo{number}{1} (\bibinfo{year}{2012}), \bibinfo{pages}{54--72}.
\newblock


\bibitem[{Le Goues} et~al\mbox{.}(2019)]%
        {cacm2019-program-repair}
\bibfield{author}{\bibinfo{person}{Claire {Le Goues}}, \bibinfo{person}{Michael Pradel}, {and} \bibinfo{person}{Abhik Roychoudhury}.} \bibinfo{year}{2019}\natexlab{}.
\newblock \showarticletitle{Automated program repair}.
\newblock \bibinfo{journal}{\emph{Commun. {ACM}}} \bibinfo{volume}{62}, \bibinfo{number}{12} (\bibinfo{year}{2019}), \bibinfo{pages}{56--65}.
\newblock
\href{https://doi.org/10.1145/3318162}{doi:\nolinkurl{10.1145/3318162}}


\bibitem[Li et~al\mbox{.}(2023)]%
        {Li2023a}
\bibfield{author}{\bibinfo{person}{Haonan Li}, \bibinfo{person}{Yu Hao}, \bibinfo{person}{Yizhuo Zhai}, {and} \bibinfo{person}{Zhiyun Qian}.} \bibinfo{year}{2023}\natexlab{}.
\newblock \bibinfo{title}{The Hitchhiker's Guide to Program Analysis: A Journey with Large Language Models}.
\newblock
\showeprint[arxiv]{2308.00245}~[cs.SE]


\bibitem[Li et~al\mbox{.}(2024)]%
        {Li2024}
\bibfield{author}{\bibinfo{person}{Ziyang Li}, \bibinfo{person}{Saikat Dutta}, {and} \bibinfo{person}{Mayur Naik}.} \bibinfo{year}{2024}\natexlab{}.
\newblock \bibinfo{title}{LLM-Assisted Static Analysis for Detecting Security Vulnerabilities}.
\newblock
\showeprint[arxiv]{2405.17238}~[cs.CR]


\bibitem[Liu et~al\mbox{.}(2018)]%
        {Liu2018a}
\bibfield{author}{\bibinfo{person}{Kui Liu}, \bibinfo{person}{Dongsun Kim}, \bibinfo{person}{Tegawend{\'e}~F Bissyand{\'e}}, \bibinfo{person}{Shin Yoo}, {and} \bibinfo{person}{Yves Le~Traon}.} \bibinfo{year}{2018}\natexlab{}.
\newblock \showarticletitle{Mining fix patterns for findbugs violations}.
\newblock \bibinfo{journal}{\emph{IEEE Transactions on Software Engineering}} (\bibinfo{year}{2018}).
\newblock


\bibitem[Liu et~al\mbox{.}(2019)]%
        {Liu2019a}
\bibfield{author}{\bibinfo{person}{Kui Liu}, \bibinfo{person}{Anil Koyuncu}, \bibinfo{person}{Dongsun Kim}, {and} \bibinfo{person}{Tegawend{\'{e}}~F. Bissyand{\'{e}}}.} \bibinfo{year}{2019}\natexlab{}.
\newblock \showarticletitle{TBar: revisiting template-based automated program repair}. In \bibinfo{booktitle}{\emph{ISSTA}}. \bibinfo{publisher}{{ACM}}, \bibinfo{pages}{31--42}.
\newblock
\href{https://doi.org/10.1145/3293882.3330577}{doi:\nolinkurl{10.1145/3293882.3330577}}


\bibitem[Liu et~al\mbox{.}(2023)]%
        {Liu2023}
\bibfield{author}{\bibinfo{person}{Yu Liu}, \bibinfo{person}{Sergey Mechtaev}, \bibinfo{person}{Pavle Suboti{\'c}}, {and} \bibinfo{person}{Abhik Roychoudhury}.} \bibinfo{year}{2023}\natexlab{}.
\newblock \showarticletitle{Program Repair Guided by Datalog-Defined Static Analysis}. In \bibinfo{booktitle}{\emph{ESEC/FSE}}. \bibinfo{pages}{1216--1228}.
\newblock


\bibitem[Long and Rinard(2016)]%
        {Long2016}
\bibfield{author}{\bibinfo{person}{Fan Long} {and} \bibinfo{person}{Martin Rinard}.} \bibinfo{year}{2016}\natexlab{}.
\newblock \showarticletitle{Automatic patch generation by learning correct code}. In \bibinfo{booktitle}{\emph{POPL}}. \bibinfo{pages}{298--312}.
\newblock


\bibitem[Lutellier et~al\mbox{.}(2020)]%
        {Lutellier2020}
\bibfield{author}{\bibinfo{person}{Thibaud Lutellier}, \bibinfo{person}{Hung~Viet Pham}, \bibinfo{person}{Lawrence Pang}, \bibinfo{person}{Yitong Li}, \bibinfo{person}{Moshi Wei}, {and} \bibinfo{person}{Lin Tan}.} \bibinfo{year}{2020}\natexlab{}.
\newblock \showarticletitle{CoCoNuT: combining context-aware neural translation models using ensemble for program repair}. In \bibinfo{booktitle}{\emph{ISSTA}}. \bibinfo{publisher}{{ACM}}, \bibinfo{pages}{101--114}.
\newblock
\href{https://doi.org/10.1145/3395363.3397369}{doi:\nolinkurl{10.1145/3395363.3397369}}


\bibitem[Marginean et~al\mbox{.}(2019)]%
        {marginean}
\bibfield{author}{\bibinfo{person}{Alexandru Marginean}, \bibinfo{person}{Johannes Bader}, \bibinfo{person}{Satish Chandra}, \bibinfo{person}{Mark Harman}, \bibinfo{person}{Yue Jia}, \bibinfo{person}{Ke Mao}, \bibinfo{person}{Alexander Mols}, {and} \bibinfo{person}{Andrew Scott}.} \bibinfo{year}{2019}\natexlab{}.
\newblock \showarticletitle{Sapfix: Automated end-to-end repair at scale}. In \bibinfo{booktitle}{\emph{2019 IEEE/ACM 41st International Conference on Software Engineering: Software Engineering in Practice (ICSE-SEIP)}}.
\newblock


\bibitem[Oh and Oh(2022)]%
        {Oh2022}
\bibfield{author}{\bibinfo{person}{Wonseok Oh} {and} \bibinfo{person}{Hakjoo Oh}.} \bibinfo{year}{2022}\natexlab{}.
\newblock \showarticletitle{PyTER: Effective Program Repair for Python Type Errors}. In \bibinfo{booktitle}{\emph{ESEC/FSE}}.
\newblock


\bibitem[Reif et~al\mbox{.}(2019)]%
        {Reif2019}
\bibfield{author}{\bibinfo{person}{Michael Reif}, \bibinfo{person}{Florian K{\"{u}}bler}, \bibinfo{person}{Michael Eichberg}, \bibinfo{person}{Dominik Helm}, {and} \bibinfo{person}{Mira Mezini}.} \bibinfo{year}{2019}\natexlab{}.
\newblock \showarticletitle{Judge: identifying, understanding, and evaluating sources of unsoundness in call graphs}. In \bibinfo{booktitle}{\emph{Proceedings of the 28th {ACM} {SIGSOFT} International Symposium on Software Testing and Analysis, {ISSTA} 2019, Beijing, China, July 15-19, 2019}}, \bibfield{editor}{\bibinfo{person}{Dongmei Zhang} {and} \bibinfo{person}{Anders M{\o}ller}} (Eds.). \bibinfo{publisher}{{ACM}}, \bibinfo{pages}{251--261}.
\newblock
\href{https://doi.org/10.1145/3293882.3330555}{doi:\nolinkurl{10.1145/3293882.3330555}}


\bibitem[Rutar et~al\mbox{.}(2004)]%
        {Rutar2004}
\bibfield{author}{\bibinfo{person}{Nick Rutar}, \bibinfo{person}{Christian~B. Almazan}, {and} \bibinfo{person}{Jeffrey~S. Foster}.} \bibinfo{year}{2004}\natexlab{}.
\newblock \showarticletitle{A Comparison of Bug Finding Tools for Java}. In \bibinfo{booktitle}{\emph{International Symposium on Software Reliability Engineering (ISSRE)}}. \bibinfo{publisher}{IEEE Computer Society}, \bibinfo{pages}{245--256}.
\newblock


\bibitem[Sakkas et~al\mbox{.}(2020)]%
        {Sakkas2020}
\bibfield{author}{\bibinfo{person}{Georgios Sakkas}, \bibinfo{person}{Madeline Endres}, \bibinfo{person}{Benjamin Cosman}, \bibinfo{person}{Westley Weimer}, {and} \bibinfo{person}{Ranjit Jhala}.} \bibinfo{year}{2020}\natexlab{}.
\newblock \showarticletitle{Type error feedback via analytic program repair}. In \bibinfo{booktitle}{\emph{Proceedings of the 41st {ACM} {SIGPLAN} International Conference on Programming Language Design and Implementation, {PLDI} 2020, London, UK, June 15-20, 2020}}, \bibfield{editor}{\bibinfo{person}{Alastair~F. Donaldson} {and} \bibinfo{person}{Emina Torlak}} (Eds.). \bibinfo{publisher}{{ACM}}, \bibinfo{pages}{16--30}.
\newblock
\href{https://doi.org/10.1145/3385412.3386005}{doi:\nolinkurl{10.1145/3385412.3386005}}


\bibitem[Silva et~al\mbox{.}(2024)]%
        {Silva2024}
\bibfield{author}{\bibinfo{person}{André Silva}, \bibinfo{person}{Sen Fang}, {and} \bibinfo{person}{Martin Monperrus}.} \bibinfo{year}{2024}\natexlab{}.
\newblock \bibinfo{title}{RepairLLaMA: Efficient Representations and Fine-Tuned Adapters for Program Repair}.
\newblock
\showeprint[arxiv]{2312.15698}~[cs.SE]


\bibitem[Staicu et~al\mbox{.}(2020)]%
        {icse2020}
\bibfield{author}{\bibinfo{person}{Cristian{-}Alexandru Staicu}, \bibinfo{person}{Martin~Toldam Torp}, \bibinfo{person}{Max Sch{\"{a}}fer}, \bibinfo{person}{Anders M{\o}ller}, {and} \bibinfo{person}{Michael Pradel}.} \bibinfo{year}{2020}\natexlab{}.
\newblock \showarticletitle{Extracting taint specifications for JavaScript libraries}. In \bibinfo{booktitle}{\emph{{ICSE} '20: 42nd International Conference on Software Engineering, Seoul, South Korea, 27 June - 19 July, 2020}}, \bibfield{editor}{\bibinfo{person}{Gregg Rothermel} {and} \bibinfo{person}{Doo{-}Hwan Bae}} (Eds.). \bibinfo{publisher}{{ACM}}, \bibinfo{pages}{198--209}.
\newblock
\href{https://doi.org/10.1145/3377811.3380390}{doi:\nolinkurl{10.1145/3377811.3380390}}


\bibitem[Tufano et~al\mbox{.}(2019)]%
        {Tufano2019}
\bibfield{author}{\bibinfo{person}{Michele Tufano}, \bibinfo{person}{Jevgenija Pantiuchina}, \bibinfo{person}{Cody Watson}, \bibinfo{person}{Gabriele Bavota}, {and} \bibinfo{person}{Denys Poshyvanyk}.} \bibinfo{year}{2019}\natexlab{}.
\newblock \showarticletitle{On learning meaningful code changes via neural machine translation}. In \bibinfo{booktitle}{\emph{ICSE}}. \bibinfo{pages}{25--36}.
\newblock
\urldef\tempurl%
\url{https://dl.acm.org/citation.cfm?id=3339509}
\showURL{%
\tempurl}


\bibitem[Utture et~al\mbox{.}(2022)]%
        {Utture2022}
\bibfield{author}{\bibinfo{person}{Akshay Utture}, \bibinfo{person}{Shuyang Liu}, \bibinfo{person}{Christian~Gram Kalhauge}, {and} \bibinfo{person}{Jens Palsberg}.} \bibinfo{year}{2022}\natexlab{}.
\newblock \showarticletitle{Striking a Balance: Pruning False-Positives from Static Call Graphs}. In \bibinfo{booktitle}{\emph{ICSE}}.
\newblock


\bibitem[Wadhwa et~al\mbox{.}(2024)]%
        {Wadhwa2024CORE:LLMs}
\bibfield{author}{\bibinfo{person}{Nalin Wadhwa}, \bibinfo{person}{Jui Pradhan}, \bibinfo{person}{Atharv Sonwane}, \bibinfo{person}{Surya~Prakash Sahu}, \bibinfo{person}{Nagarajan Natarajan}, \bibinfo{person}{Aditya Kanade}, \bibinfo{person}{Suresh Parthasarathy}, {and} \bibinfo{person}{Sriram Rajamani}.} \bibinfo{year}{2024}\natexlab{}.
\newblock \showarticletitle{{CORE: Resolving Code Quality Issues using LLMs}}.
\newblock \bibinfo{journal}{\emph{Proceedings of the ACM on Software Engineering}} \bibinfo{volume}{1}, \bibinfo{number}{FSE} (\bibinfo{date}{7} \bibinfo{year}{2024}), \bibinfo{pages}{789--811}.
\newblock
\showISSN{2994-970X}
\href{https://doi.org/10.1145/3643762}{doi:\nolinkurl{10.1145/3643762}}


\bibitem[Wang et~al\mbox{.}(2021)]%
        {Wang2021c}
\bibfield{author}{\bibinfo{person}{Tongjie Wang}, \bibinfo{person}{Yaroslav Golubev}, \bibinfo{person}{Oleg Smirnov}, \bibinfo{person}{Jiawei Li}, \bibinfo{person}{Timofey Bryksin}, {and} \bibinfo{person}{Iftekhar Ahmed}.} \bibinfo{year}{2021}\natexlab{}.
\newblock \showarticletitle{{PyNose}: A Test Smell Detector For {P}ython}. In \bibinfo{booktitle}{\emph{ASE}}.
\newblock


\bibitem[Wang et~al\mbox{.}(2024)]%
        {Wang2024a}
\bibfield{author}{\bibinfo{person}{Xingyao Wang}, \bibinfo{person}{Boxuan Li}, \bibinfo{person}{Yufan Song}, \bibinfo{person}{Frank~F. Xu}, \bibinfo{person}{Xiangru Tang}, \bibinfo{person}{Mingchen Zhuge}, \bibinfo{person}{Jiayi Pan}, \bibinfo{person}{Yueqi Song}, \bibinfo{person}{Bowen Li}, \bibinfo{person}{Jaskirat Singh}, \bibinfo{person}{Hoang~H. Tran}, \bibinfo{person}{Fuqiang Li}, \bibinfo{person}{Ren Ma}, \bibinfo{person}{Mingzhang Zheng}, \bibinfo{person}{Bill Qian}, \bibinfo{person}{Yanjun Shao}, \bibinfo{person}{Niklas Muennighoff}, \bibinfo{person}{Yizhe Zhang}, \bibinfo{person}{Binyuan Hui}, \bibinfo{person}{Junyang Lin}, \bibinfo{person}{Robert Brennan}, \bibinfo{person}{Hao Peng}, \bibinfo{person}{Heng Ji}, {and} \bibinfo{person}{Graham Neubig}.} \bibinfo{year}{2024}\natexlab{}.
\newblock \bibinfo{title}{OpenHands: An Open Platform for AI Software Developers as Generalist Agents}.
\newblock
\showeprint[arxiv]{2407.16741}~[cs.SE]
\urldef\tempurl%
\url{https://arxiv.org/abs/2407.16741}
\showURL{%
\tempurl}


\bibitem[Wen et~al\mbox{.}(2024)]%
        {Wen2024}
\bibfield{author}{\bibinfo{person}{Cheng Wen}, \bibinfo{person}{Yuandao Cai}, \bibinfo{person}{Bin Zhang}, \bibinfo{person}{Jie Su}, \bibinfo{person}{Zhiwu Xu}, \bibinfo{person}{Dugang Liu}, \bibinfo{person}{Shengchao Qin}, \bibinfo{person}{Zhong Ming}, {and} \bibinfo{person}{Cong Tian}.} \bibinfo{year}{2024}\natexlab{}.
\newblock \showarticletitle{Automatically Inspecting Thousands of Static Bug Warnings with Large Language Model: How Far Are We?}
\newblock \bibinfo{journal}{\emph{ACM Transactions on Knowledge Discovery from Data}} (\bibinfo{year}{2024}).
\newblock


\bibitem[Wu et~al\mbox{.}(2024)]%
        {Wu2024iSMELL}
\bibfield{author}{\bibinfo{person}{Di Wu}, \bibinfo{person}{Fangwen Mu}, \bibinfo{person}{Lin Shi}, \bibinfo{person}{Zhaoqiang Guo}, \bibinfo{person}{Kui Liu}, \bibinfo{person}{Weiguang Zhuang}, \bibinfo{person}{Yuqi Zhong}, {and} \bibinfo{person}{Li Zhang}.} \bibinfo{year}{2024}\natexlab{}.
\newblock \showarticletitle{iSMELL: Assembling LLMs with Expert Toolsets for Code Smell Detection and Refactoring}. In \bibinfo{booktitle}{\emph{Proceedings of the 39th IEEE/ACM International Conference on Automated Software Engineering}} (Sacramento, CA, USA) \emph{(\bibinfo{series}{ASE '24})}. \bibinfo{publisher}{Association for Computing Machinery}, \bibinfo{address}{New York, NY, USA}, \bibinfo{pages}{1345–1357}.
\newblock
\showISBNx{9798400712487}
\href{https://doi.org/10.1145/3691620.3695508}{doi:\nolinkurl{10.1145/3691620.3695508}}


\bibitem[Xia and Zhang(2024)]%
        {Xia2024a}
\bibfield{author}{\bibinfo{person}{Chunqiu~Steven Xia} {and} \bibinfo{person}{Lingming Zhang}.} \bibinfo{year}{2024}\natexlab{}.
\newblock \showarticletitle{Automated Program Repair via Conversation: Fixing 162 out of 337 Bugs for {\textdollar}0.42 Each using ChatGPT}. In \bibinfo{booktitle}{\emph{Proceedings of the 33rd {ACM} {SIGSOFT} International Symposium on Software Testing and Analysis, {ISSTA} 2024, Vienna, Austria, September 16-20, 2024}}, \bibfield{editor}{\bibinfo{person}{Maria Christakis} {and} \bibinfo{person}{Michael Pradel}} (Eds.). \bibinfo{publisher}{{ACM}}, \bibinfo{pages}{819--831}.
\newblock
\href{https://doi.org/10.1145/3650212.3680323}{doi:\nolinkurl{10.1145/3650212.3680323}}


\bibitem[Ye et~al\mbox{.}(2022)]%
        {Ye2022a}
\bibfield{author}{\bibinfo{person}{He Ye}, \bibinfo{person}{Matias Martinez}, {and} \bibinfo{person}{Martin Monperrus}.} \bibinfo{year}{2022}\natexlab{}.
\newblock \showarticletitle{Neural Program Repair with Execution-based Backpropagation}. In \bibinfo{booktitle}{\emph{ICSE}}.
\newblock


\bibitem[Ye and Monperrus(2024)]%
        {Ye2024}
\bibfield{author}{\bibinfo{person}{He Ye} {and} \bibinfo{person}{Martin Monperrus}.} \bibinfo{year}{2024}\natexlab{}.
\newblock \showarticletitle{ITER: Iterative Neural Repair for Multi-Location Patches}. In \bibinfo{booktitle}{\emph{ICSE}}.
\newblock


\bibitem[Youn et~al\mbox{.}(2023)]%
        {Youn2023}
\bibfield{author}{\bibinfo{person}{Dongjun Youn}, \bibinfo{person}{Sungho Lee}, {and} \bibinfo{person}{Sukyoung Ryu}.} \bibinfo{year}{2023}\natexlab{}.
\newblock \showarticletitle{Declarative static analysis for multilingual programs using CodeQL}.
\newblock \bibinfo{journal}{\emph{Softw. Pract. Exp.}} \bibinfo{volume}{53}, \bibinfo{number}{7} (\bibinfo{year}{2023}), \bibinfo{pages}{1472--1495}.
\newblock
\href{https://doi.org/10.1002/spe.3199}{doi:\nolinkurl{10.1002/spe.3199}}


\bibitem[Zhao et~al\mbox{.}(2018)]%
        {Zhao2018a}
\bibfield{author}{\bibinfo{person}{Jinman Zhao}, \bibinfo{person}{Aws Albarghouthi}, \bibinfo{person}{Vaibhav Rastogi}, \bibinfo{person}{Somesh Jha}, {and} \bibinfo{person}{Damien Octeau}.} \bibinfo{year}{2018}\natexlab{}.
\newblock \showarticletitle{Neural-augmented static analysis of Android communication}. In \bibinfo{booktitle}{\emph{Proceedings of the 2018 {ACM} Joint Meeting on European Software Engineering Conference and Symposium on the Foundations of Software Engineering, {ESEC/SIGSOFT} {FSE} 2018, Lake Buena Vista, FL, USA, November 04-09, 2018}}. \bibinfo{pages}{342--353}.
\newblock
\href{https://doi.org/10.1145/3236024.3236066}{doi:\nolinkurl{10.1145/3236024.3236066}}


\bibitem[Zheng et~al\mbox{.}(2021)]%
        {Zheng2021}
\bibfield{author}{\bibinfo{person}{Yunhui Zheng}, \bibinfo{person}{Saurabh Pujar}, \bibinfo{person}{Burn~L. Lewis}, \bibinfo{person}{Luca Buratti}, \bibinfo{person}{Edward~A. Epstein}, \bibinfo{person}{Bo Yang}, \bibinfo{person}{Jim Laredo}, \bibinfo{person}{Alessandro Morari}, {and} \bibinfo{person}{Zhong Su}.} \bibinfo{year}{2021}\natexlab{}.
\newblock \showarticletitle{{D2A:} {A} Dataset Built for AI-Based Vulnerability Detection Methods Using Differential Analysis}. In \bibinfo{booktitle}{\emph{43rd {IEEE/ACM} International Conference on Software Engineering: Software Engineering in Practice, {ICSE} {(SEIP)} 2021, Madrid, Spain, May 25-28, 2021}}. \bibinfo{publisher}{{IEEE}}, \bibinfo{pages}{111--120}.
\newblock
\href{https://doi.org/10.1109/ICSE-SEIP52600.2021.00020}{doi:\nolinkurl{10.1109/ICSE-SEIP52600.2021.00020}}


\end{thebibliography}

\end{document}